\documentclass{aa}
\usepackage[varg]{txfonts}
\usepackage{natbib}
\usepackage{graphicx}
\usepackage{amsmath}
\bibpunct{(}{)}{;}{a}{}{,}

\begin{document}

\title{Analytical formulation of lunar cratering asymmetries}

\author{Nan Wang         
    \and Ji-Lin Zhou     
    }

\institute{School of Astronomy and Space Science and Key Laboratory of Modern Astronomy and Astrophysics in Ministry of Education, Nanjing University, Nanjing 210046, China  \\
        \email{zhoujl@nju.edu.cn}}

\date{Received / Accepted }

\abstract
{The cratering asymmetry of a bombarded satellite is related to both its orbit and impactors. The inner solar system impactor populations, that is, the main-belt asteroids (MBAs) and the near-Earth objects (NEOs), have dominated during the late heavy bombardment (LHB) and ever since, respectively.}
{We formulate the lunar cratering distribution and verify the cratering asymmetries generated by the MBAs as well as the NEOs.}
{Based on a planar model that excludes the terrestrial and lunar gravitations on the impactors and assuming the impactor encounter speed with Earth $v_{\rm{enc}}$ is higher than the lunar orbital speed $v_{\rm{M}}$, we rigorously integrated the lunar cratering distribution, and derived its approximation to the first order of $v_{\rm{M}}/v_{\rm{enc}}$. Numerical simulations of lunar bombardment by the MBAs during the LHB were performed with an Earth-Moon distance $a_{\rm{M}}$ = 20--60 Earth radii in five cases.}
{The analytical model directly proves the existence of a leading/trailing asymmetry and the absence of near/far asymmetry. The approximate form of the leading/trailing asymmetry is $(1 + A_1 \cos\beta)$, which decreases as the apex distance $\beta$ increases.
The numerical simulations show evidence of a pole/equator asymmetry as well as the leading/trailing asymmetry, and the former is empirically described as $(1 + A_2 \cos2\varphi)$, which decreases as the latitude modulus $|\varphi|$ increases.
The amplitudes $A_{1,2}$ are reliable measurements of asymmetries. Our analysis explicitly indicates the quantitative relations between cratering distribution and bombardment conditions (impactor properties and the lunar orbital status) like $A_1 \propto v_{\rm{M}}/v_{\rm{enc}}$, resulting in a method for reproducing the bombardment conditions through measuring the asymmetry.
Mutual confirmation between analytical model and numerical simulations is found in terms of the cratering distribution and its variation with $a_{\rm{M}}$. Estimates of $A_1$ for crater density distributions generated by the MBAs and the NEOs are 0.101--0.159 and 0.117, respectively.}
{}


\authorrunning{Nan Wang \& Ji-Lin Zhou}
\maketitle

\section{Introduction}   \label{sec-intro}

The nonuniformity of the cratering distribution on a satellite or a planet when it comes under bombardment is called cratering asymmetry. There are three types. The first is the leading/trailing asymmetry, which is due to the synchronous rotation of a satellite. When synchronously locked, a satellite's velocity always points to its leading side, so that this hemisphere tends to gain a higher impact probability, higher impact speed and more normal impacts. Because the crater diameter depends on impact speed and incidence angle \citetext{e.g., \citealp{LeFeuvre2011}}, this can also lead to an enhanced crater size on the leading side. This is the so-called apex/antapex effect \citep{Zahnle2001, LeFeuvre2011}. \citet{Zahnle1998, Zahnle2001} and \citet{Levison2000} investigated this effect on the giant planets and their satellites in detail. This work focuses on the Moon. Its leading/trailing asymmetry has been proposed by theoretical works \citep{Zahnle1998, LeFeuvre2011}, confirmed by numerical simulations \citep{Gallant2009, Ito2010}, and directly verified by seismic observations \citep{Kawamura2011, Oberst2012} and observations of rayed craters \citep{Morota2003}.
The second type, the pole/equator asymmetry (or latitudinal asymmetry), which is an enhancement of low-latitude impacts compared to high-latitude regions resulting from the concentration of low-inclination projectiles, has often been reported as well \citep{LeFeuvre2008, LeFeuvre2011, Gallant2009, Ito2010}.
The third type is the near/far asymmetry, which is most uncertain. It has been proposed that Earth focuses the projectiles onto the near side of the Moon as a gravitational lens and thus the craters there are enhanced \citep{Wiesel1971}, or that Earth is an obstacle in the projectiles' trajectories so that the near side is shielded from impacts \citep{Bandermann1973}. Contradicting and limited conclusions have not led to consensus.

The cratering asymmetry depends on not only the lunar orbit, but also on the impactor population. After analyzing the size distributions of craters, \citet{Strom2005, Strom2015} suggested that there are two impactor populations in the inner solar system: the main-belt asteroids (MBAs), which dominated during the late heavy bombardment (LHB) $\sim$ 3.9 Gya, and the near-Earth objects (NEOs), which have dominated since about 3.8--3.7 Gya. The two populations are different in their orbital and size distributions, fluxes, and origins, which means that there is no reason to expect the cratering asymmetries generated by them to be the same.

We note the MBAs referred to here are the asteroids that occupied the region where the current main belt is during the LHB. Although their semi-major axes $a_{\rm{p}}$ = 2.0--3.5 AU were the same as today, their eccentricities and inclinations were greatly excited by migrating giant planets \citep{Gomes2005}. Instead, their size distribution has changed little after the first $\sim 100$ Myr \citep{Bottke2005}. Several surveys of the current main belt have reported power-law breaks of its size distribution. The Sloan Digital Sky Survey \citetext{SDSS; \citealp{Ivezic2001}} distinguished two types of asteroids with different albedos (0.04 and 0.16 for blue and red asteroids) and found that for the whole sample throughout the whole belt, the cumulative size distribution has a slope $\alpha_{\rm{p}} = 1.3$ for diameters $d_{\rm{p}}$ = 0.4--5 km and $3$ for 5--40 km. \citet{Parker2008} also confirmed this conclusion based on the SDSS moving object catalog 4, including $\sim$88000 objects. The first \citep{Yoshida2003} and second \citep{Yoshida2007} runs of the Sub-km Main Belt Asteroid Survey (SMBAS), which each found 861 MBAs in R band and 1001 MBAs in both B and R bands, reported $\alpha_{\rm{p}} = 1.19\pm0.02$ for $d_{\rm{p}}$ = 0.5--1 km and $\alpha_{\rm{p}} = 1.29\pm0.02$ for $d_{\rm{p}}$ = 0.6--1 km, claimed to be consistent with SDSS for MBAs smaller than 5 km. The size distribution of craters formed by MBAs also has a complex shape, whose cumulative slope is $\alpha_{\rm{c}}$ = 1.2 for crater diameters $d_{\rm{c}} \lesssim 50$ km, 2 for 50--100 km, and 3 for 100--300 km \citep{Strom2015}.

The NEO population is relatively well understood and suggested to have been in steady state for the past $\sim$ 3 Gyr, constantly resupplied mainly from the main belt \citep{Bottke2002}. \citet{Bottke2002} derived the debiased orbital and size distributions of NEOs by fitting a model population to the known NEOs: the orbits are constrained in a range $a_{\rm{p}}$ = 0.5--2.8 AU, $e_{\rm{p}} < 0.8$, $i_{\rm{p}} < 35\degr$, the perihelion distance $q_{\rm{p}} < 1.3$ AU, and the aphelion distance $Q_{\rm{p}} > 0.983$ AU; the size distribution is characterized by a single slope $\alpha_{\rm{p}} = 1.75 \pm 0.1$ for $d_{\rm{p}}$ = 0.2--4 km. The corresponding slope for the crater size distribution indicated by \citet{Strom2015} is $\alpha_{\rm{c}}$ = 2 for $d_{\rm{c}}$ = 0.02--100 km.
A few recent works have investigated the cratering asymmetry generated by NEOs based on the debiased NEO model described in \citet{Bottke2002}. \citet{Gallant2009} ran $N$-body simulations and determined an apex/antapex ratio (ratio of the crater density at the apex to antapex) of $1.28 \pm 0.01$,
a polar deficiency of $\sim10\%$,
and the absence of a near/far asymmetry. In a similar work but with a different numerical model, \citet{Ito2010} found a leading/trailing hemispherical ratio of $1.32 \pm 0.01$ and also a polar deficiency of $\sim10\%$. \citet{LeFeuvre2008} analytically predicted the lunar pole/equator ratio to be 0.90. \citet{LeFeuvre2011} followed and semi-analytically estimated that involving both longitudinal and latitudinal asymmetries, the lunar crater density was minimized at ($90\degr$ E, $\pm65\degr$ N) and maximized at the apex with deviations of about 25\% with respect to the average for the current Earth-Moon distance, resulting in an apex/antapex ratio of 1.37 and a pole/equator ratio of 0.80. From observations, \citet{Morota2003} reported an apex/anntapex ratio of $\sim1.5$ for 222 rayed craters with $d_{\rm{c}} > 5$ km on lunar highlands, formed by NEOs in the past $\sim1.1$ Gyr.

However, according to \citet{Strom2015}, for craters with diameters larger than 10 km, those that the MBAs formed exceed the other population by more than an order of magnitude. This underlines the need for taking the craters and the cratering asymmetry that the MBAs contribute to into account, and thus requires the awareness of the lunar orbit during the LHB, namely, the dominance of MBAs. The Earth-Moon system evolved drastically in the early history. As a general trend, it is accepted that the Moon has been receding from Earth because of tidal dissipation, but a wide spectrum of opinions exists for the details of this picture. After the timescale problem was raised \citep{Gerstenkorn1955, MacDonald1964, Goldreich1966, Lambeck1977, Touma1994} and then solved through ocean models \citep{Hansen1982, Webb1982, Ross1989, Kagan1994, Kagan1997, Bills1999}, it is known today that the tidal dissipation factor of Earth must have been much larger in the distant past, or in other words, the tidal friction was much weaker than today \citetext{e.g., \citealp{Bills1999}}. Because tidal friction depends on not only the lunar orbit but also the terrestrial ocean shape, which is associated with continental drift, the exact history of Earth's tidal dissipation factor and hence the early evolution of the lunar orbit cannot be confirmed.

Still, we list some efforts of ocean modelers here. \citet{Hansen1982} was the first to introduce Laplace's tidal equations in calculations of oceanic tidal torque. He modeled four cases, combinations of two idealized continentalities and two frictional resistance coefficients, and found the Earth-Moon distances at 4.5 Gya ranging from 38 to 53 $R_{\oplus}$, leading to nearly the same values at 3.9 Gya when the LHB occurred. \citet{Webb1980, Webb1982} developed an average ocean model and obtained the dates of the Gerstenkorn event as 3.9 and 5.3 Gya with and without solid Earth dissipation included, respectively. According to these two results, the Earth-Moon distance at 3.9 Gya should be no larger than 25 $R_{\oplus}$ or about 42 $R_{\oplus}$. The author claimed, however, that the results should only be taken qualitatively. \citet{Ross1989} simulated a coupled thermal-dynamical evolution of the Earth-Moon system based on equilibrium ocean model, resulting in $a_{\rm{M}}$ being 47 $R_{\oplus}$ and $e_{\rm{M}}$ being 0.04 at 3.9 Gya when the evolution timescale and parameters such as final $a_{\rm{M}}$ and $e_{\rm{M}}$ were required to be realistic. \citet{Kagan1994} described a stochastic model that considered fluctuating effects of the continental drift. Their two-mode resonance approximations gave rise to evolutions of tidal energy dissipation that were consistent with global paleotide models. By adopting a reproduced tidal evolution with a timescale as close as possible to the realistic one, the Earth-Moon distance is estimated to be about 47 $R_{\oplus}$ at 3.9 Gya. This timescale can vary by billions of years, of course, depending on the values of the resonance lifetime.

The uncertain lunar orbital status during the LHB means that a reliable relationship between the lunar orbit and the cratering asymmetry is needed. If the influence of the former on the latter were significant and an incorrect condition of the Earth-Moon system were assumed, the estimated cratering asymmetry would be also incorrect. Conversely, this also implies that the early history of the Earth-Moon system could be inferred from the observed crater record if the portion of craters that formed during the LHB were selected. The influence of the lunar orbit and of the impactor population on the cratering asymmetry have not been sufficiently studied so far. Numerical simulations can only offer an empirical estimation based on a limited number of cases: \citet{Zahnle2001}, who simulated impacts of ecliptic comets on giant planet satellites, combined the results of case $\alpha_{\rm{p}} = 2.0$ and case $\alpha_{\rm{p}} = 2.5$ with the satellite orbital speed $v_{\rm{orb}} = 10.9$ km~s$^{-1}$ and the impactor speed at infinity in the rest frame of the planet fixed at $v_{\infty} \approx 5$ km~s$^{-1}$, and thus derived a semi-empirical description of the crater density
\begin{equation}    \label{eq-Zahnle2001}
    N_{\rm{c}} \propto (1 + \frac{v_{\rm{orb}}}{\sqrt{2 v_{\rm{orb}}^2+v_{\infty}^2}} \cos\beta)^{2.0+(1.4/3)\alpha_{\rm{p}}},
\end{equation}
where $\beta$ is the angular distance from the apex; using the debiased NEO model, \citet{Gallant2009} ran simulations with Earth-Moon distances of $a_{\rm{M}}$ = 50, 38, 30, 20, and 10 $R_{\oplus}$, and confirmed the negative correlation between $a_{\rm{M}}$ and leading/trailing asymmetry degree. A semi-analytical method capable of producing cratering of the Moon caused by current asteroids and comets in the inner solar system was proposed by \citet{LeFeuvre2011}, who suggested a fit relation
\begin{equation}    \label{eq-LF11}
    N_{\rm{c}}({\rm{apex}}) / N_{\rm{c}}({\rm{antapex}}) = 1.12 {\rm{e}}^{-0.0529 (a_{\rm{M}}/R_{\oplus})} + 1.32,
\end{equation}
which is valid between 20 and 60 $R_{\oplus}$.

In this paper, we first derive the formulated lunar cratering distribution through rigorous integration based on a planar model in Sect. \ref{sec-analy}, where the leading/trailing asymmetry is unambiguously confirmed. Then we show the numerical simulations of the Moon under bombardment of MBAs during the LHB over various Earth-Moon distances in Sect. \ref{sec-simul}. The simulation results are compared to our analytical predictions, and the cratering distribution of coupled asymmetries is described. Last, Sect. \ref{sec-discu} compares NEOs and MBAs and shows the consistence of our analytical model with related works.

\section{Analytical deduction}   \label{sec-analy}

This section shows how we analytically deduced formulation series describing the spatial distributions of impact speed, incidence angle, normal speed, crater diameter, impact density, and crater density on the lunar surface under bombardment, based on a planar model. These formulations directly prove the existence of the leading/trailing cratering asymmetry and the identity of the near and far sides. The amplitude of the leading/trailing asymmetry is proportional to the ratio of the impactor encounter speed to the lunar orbital speed, implying that it might be possible to derive the lunar orbit or the impactor population from crater record.

\subsection{Assumptions and precondition}   \label{subsec-assum}

Our model includes the Sun, Earth, the Moon, and the impactors (asteroids and/or comets) in the inner solar system, which are assumed to be located farther away from the Sun than Earth. The main assumptions we adopt are listed below.
\begin{enumerate}
  \item The orbits of Earth, the Moon, and the impactors as well as the lunar equator are coplanar.
  \item The orbits of Earth and the Moon are circular.
  \item The gravitation on the impactors due to Earth and the Moon are ignored.
  \item Earth is treated as a particle.
\end{enumerate}
The first assumption does not allow describing latitudinal cratering variation, which is investigated in our numerical simulations. It helps avoiding the coupling of two cratering asymmetries and concluding a pure leading/trailing asymmetry formulation. The Moon's geometrical libration is neglected by the first two assumptions. The third partly means that the acceleration of the impactor velocity is not accounted for, which is valid on condition that the Earth-Moon distance is larger than $\sim 17$ $R_{\oplus}$, according to \citet{LeFeuvre2011}. It also means the gravitational lensing by Earth is excluded, and moreover, the last assumption excludes the Earth's shielding for the Moon. This causes the symmetry between the near and far sides.

A precondition for the cratering asymmetry is that the Moon must be synchronously rotating, because otherwise any longitudinal variation would vanish as the lunar hemisphere facing the Earth changes constantly. According to \citet{Zhou2008}, the timescale for the Moon to reach the 1:1 spin-orbit resonance can be estimated by
\begin{equation}
    \tau_{\rm{syn}} \sim (\frac{R_{\rm{M}}}{a_{\rm{M}}})^2 \tau_{\rm{tide}} = \frac{4 Q_{\rm{M}}'}{63 n_{\rm{M}}} \frac{m_{\rm{M}}}{m_{\oplus}} (\frac{a_{\rm{M}}}{R_{\rm{M}}})^3,
\end{equation}
where $\tau_{\rm{tide}}$ is the tidal circularization timescale, $m_{\oplus}$ is the Earth mass, $n_{\rm{M}}$, $a_{\rm{M}}$, $R_{\rm{M}}$, $m_{\rm{M}}$, and $Q_{\rm{M}}'$ are the mean motion, semi-major axis, radius, mass, and effective tidal dissipation factor of the Moon.
A measure of this timescale is $\tau_{\rm{syn}} \lesssim 10^2$ years for $a_{\rm{M}} = 10$ $R_{\oplus}$, several orders shorter than the timescale of the tidal evolution. This precondition is therefore expected to hold for the main lifetime of the Moon, including during the LHB.

On condition that the Moon rotates synchronously, its hemispheres can be defined. The near side is defined as the hemisphere that always faces Earth, and the far side is its opposite. The leading side is defined as the hemisphere that the lunar orbital velocity points to, and the trailing side is its opposite. The centers of the leading and trailing sides are the apex and the antapex points.

\subsection{Encounter geometry}   \label{subsec-enc}

\begin{figure}
    \resizebox{\hsize}{!}{\includegraphics{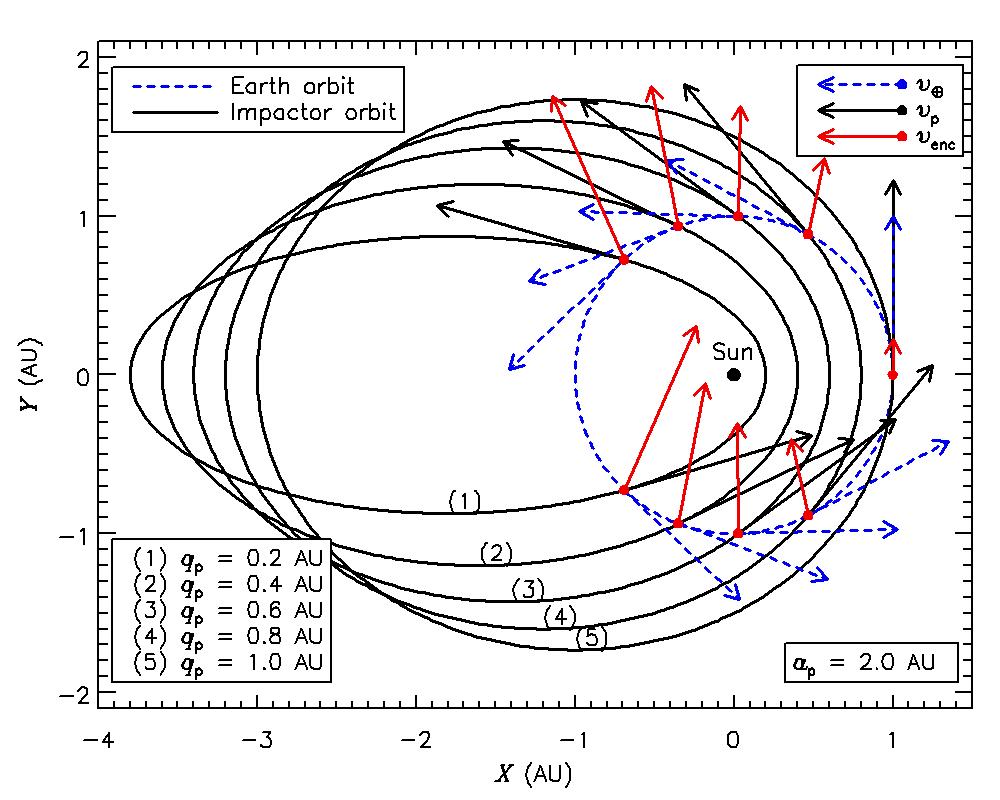}}
    \caption{Encounter geometry for an impactor orbit with $a_{\rm{p}}$ = 2.0 AU and $q_{\rm{p}}$ = 0.2--1.0 AU. Where the Earth orbit (blue dashed circle) intersects the impactor orbit (black solid ellipse), the angle between $\vec{v}_{\oplus}$ (blue dashed arrow) and $\vec{v}_{\rm{p}}$ (black solid arrow) decreases as $q_{\rm{p}}$ increases, and thus $\vec{v}_{\rm{enc}}$ (red solid arrow) shrinks.}
    \label{fig-EncGeo}
\end{figure}

\begin{figure}
    \resizebox{\hsize}{!}{\includegraphics{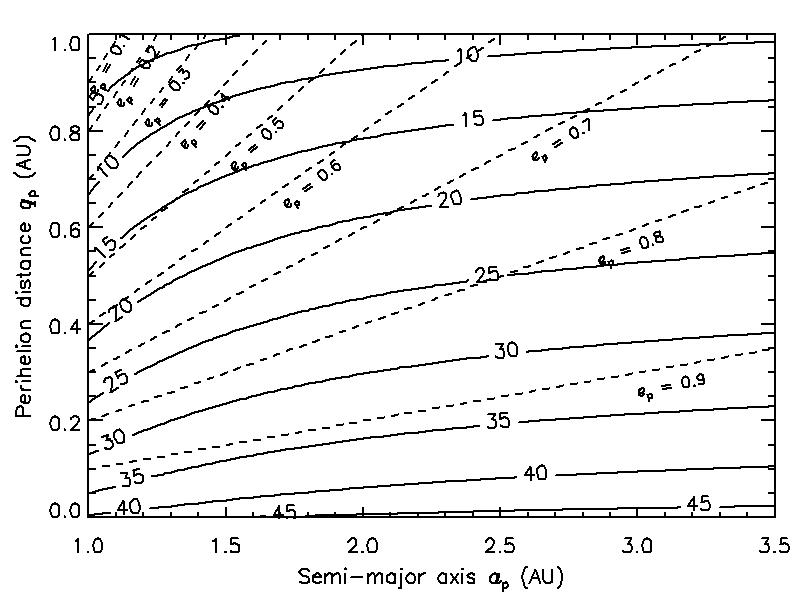}}
    \caption{Contour map of the encounter speed $v_{\rm{enc}}$ (km~s$^{-1}$) as a function of $a_{\rm{p}}$ and $q_{\rm{p}}$.}
    \label{fig-v_enc}
\end{figure}

We first consider the encounter condition before exhibiting the details of the impact in the following subsections. In the heliocentric frame, Earth orbits the Sun circularly at 1 AU, the Moon is ignored at this pre-encounter stage, and the impactors are all massless particles with semi-major axes $a_{\rm{p}} > 1$ AU and eccentricities $0 < e_{\rm{p}} < 1$. If the orbit of an impactor intersects that of Earth on the ecliptic, meaning that its perihelion distance $q_{\rm{p}} \le 1 \rm{AU}$, an encounter is considered to occur at the mutual node. The relative velocity between the impactor and Earth when they encounter is the encounter velocity $\vec{v}_{\rm{enc}}$ = $\vec{v}_{\rm{p}}$ - $\vec{v}_{\oplus}$, where $\vec{v}_{\rm{p}}$ and $\vec{v}_{\oplus}$ are orbital velocities of the impactor and Earth. Without loss of generality, the impactor's and Earth's arguments of perihelion are both assumed to be $0\degr$, so that their velocities in heliocentric ecliptic coordinates are
\begin{align}
  \vec{v}_{\rm{p}} &= \sqrt{\frac{G M_{\odot}}{a_{\rm{p}}(1-e_{\rm{p}}^2)}} (-\sin f_{\rm{p}}, e_{\rm{p}} + \cos f_{\rm{p}}),    \\
  \vec{v}_{\oplus} &= \sqrt{\frac{G M_{\odot}}{a_{\oplus}}} (-\sin f_{\oplus}, \cos f_{\oplus}),
\end{align}
where $G$ is the gravitational constant, $M_{\odot}$ is the solar mass, $f_{\rm{p}}$ and $f_{\oplus}$ are true anomalies, and the terrestrial semi-major axis is $a_{\oplus} = 1$ AU. The mutual node is where $f_{\rm{p}} = f_{\oplus} = f_{\rm{enc}}$, making the impactor's heliocentric distance equal to $a_{\oplus}$, that is,
\begin{equation}
    \cos f_{\rm{enc}} = \frac{a_{\rm{p}}(1 - e_{\rm{p}}^2)}{a_{\oplus} e_{\rm{p}}} - \frac{1}{e_{\rm{p}}}.
\end{equation}
Thus, the encounter velocity depends on $a_{\rm{p}}$ and $e_{\rm{p}}$ (or $q_{\rm{p}}$):
\begin{equation}    \label{eq-v_enc}
    \vec{v}_{\rm{enc}}(a_{\rm{p}}, e_{\rm{p}}) = \vec{v}_{\rm{p}}(a_{\rm{p}}, e_{\rm{p}}, f_{\rm{enc}}(a_{\rm{p}}, e_{\rm{p}})) - \vec{v}_{\oplus}(f_{\rm{enc}}(a_{\rm{p}}, e_{\rm{p}})).
\end{equation}

For a fixed $a_{\rm{p}}$, as shown in Fig. \ref{fig-EncGeo}, the encounter speed $v_{\rm{enc}}$ is minimized when $q_{\rm{p}} = 1$ AU. As $q_{\rm{p}}$ decreases ($e_{\rm{p}}$ increases), even though the magnitudes of the two orbital velocities $\vec{v}_{\rm{p}}$ and $\vec{v}_{\oplus}$ in the moment of encounter are invariant, the angle between them expands, and thus the encounter speed $v_{\rm{enc}}$ increases. On the other hand, if $q_{\rm{p}}$ is fixed, larger $a_{\rm{p}}$ can also lead to higher $v_{\rm{enc}}$, because the impactor orbital speed in the moment of encounter
\begin{equation}
    v_{\rm{p}} = \sqrt{G M_{\odot} (\frac{2}{a_{\oplus}} - \frac{1}{a_{\rm{p}}})}
\end{equation}
is greater. As shown in Fig. \ref{fig-v_enc}, $v_{\rm{enc}}$ increases rightward ($a_{\rm{p}}$ increases) and downward ($q_{\rm{p}}$ decreases). Given a population of impactors, its typical $v_{\rm{enc}}$ is determined by its orbital distribution and can affect the cratering distribution. Therefore, different impactor populations represent different cratering asymmetries with a given target, providing a method for deriving impactor properties form the observed crater record.

\subsection{Impact geometry}   \label{subsec-imp}

\begin{figure*}
    \centering
    \includegraphics[width=17cm]{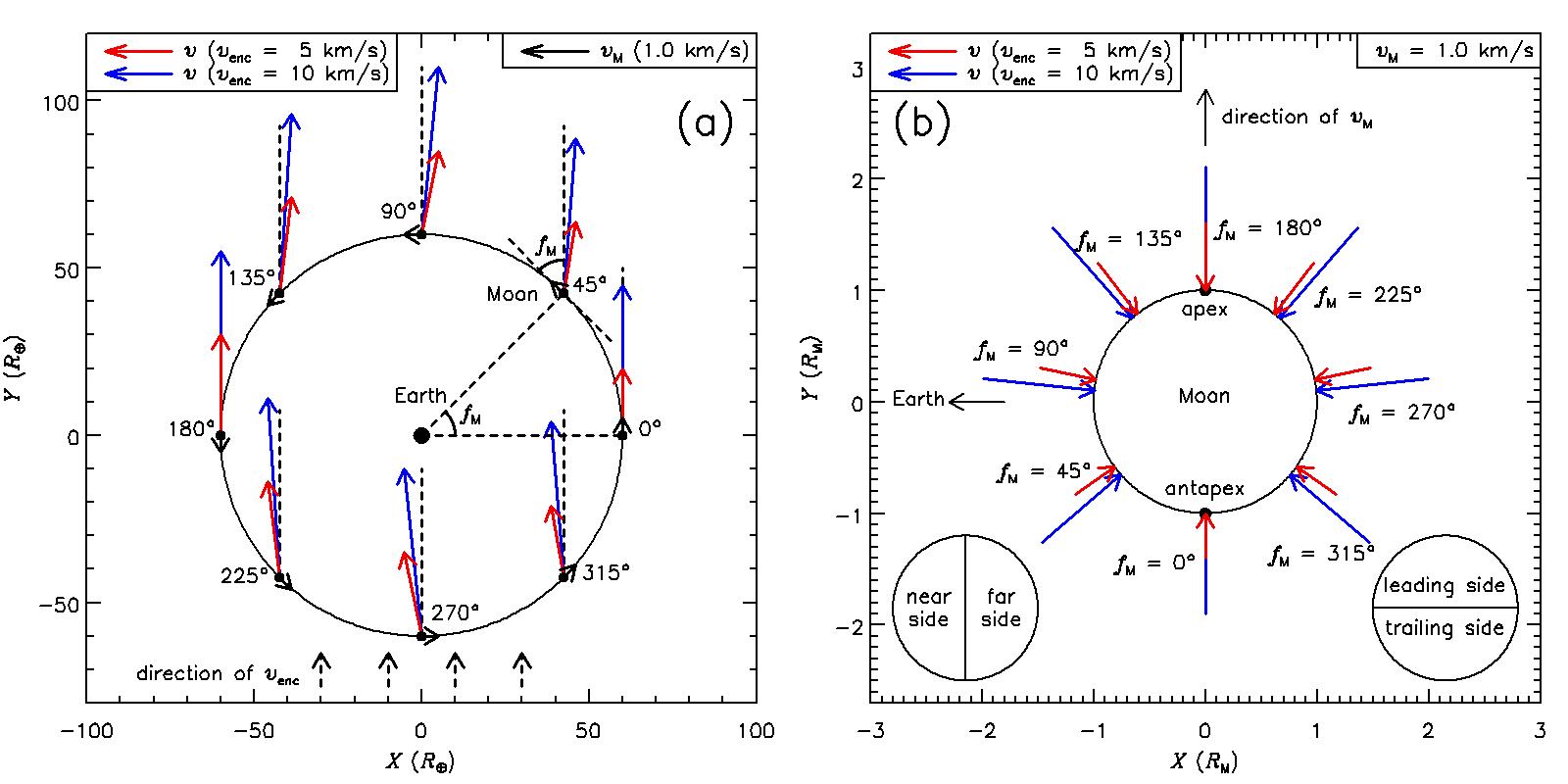}
    \caption{Impact geometry seen in the rest frame of Earth (a) and Moon (b). a) The Moon is assumed to be where $a_{\rm{M}}$ = 60 $R_{\oplus}$, with $v_{\rm{M}}$ = 1.0 km~s$^{-1}$. The impactors, distributed extensively enough to cover the lunar orbit (black circle), are in the common direction of $\vec{v}_{\rm{enc}}$ (black dashed arrow). Where $f_{\rm{M}} = 0\degr, 45\degr, \dots, 315\degr$, the lunar velocity $\vec{v}_{\rm{M}}$ (black solid arrow) and the impact velocity $\vec{v}$ for $v_{\rm{enc}}$ = 5 or 10 km~s$^{-1}$ (red or blue solid arrows) are all plotted to the same scale. b) Under the same conditions, $\vec{v}$ is plotted pointing at the normal impact points on the lunar surface.}
    \label{fig-ImpGeo}
\end{figure*}

\begin{figure}
    \resizebox{\hsize}{!}{\includegraphics{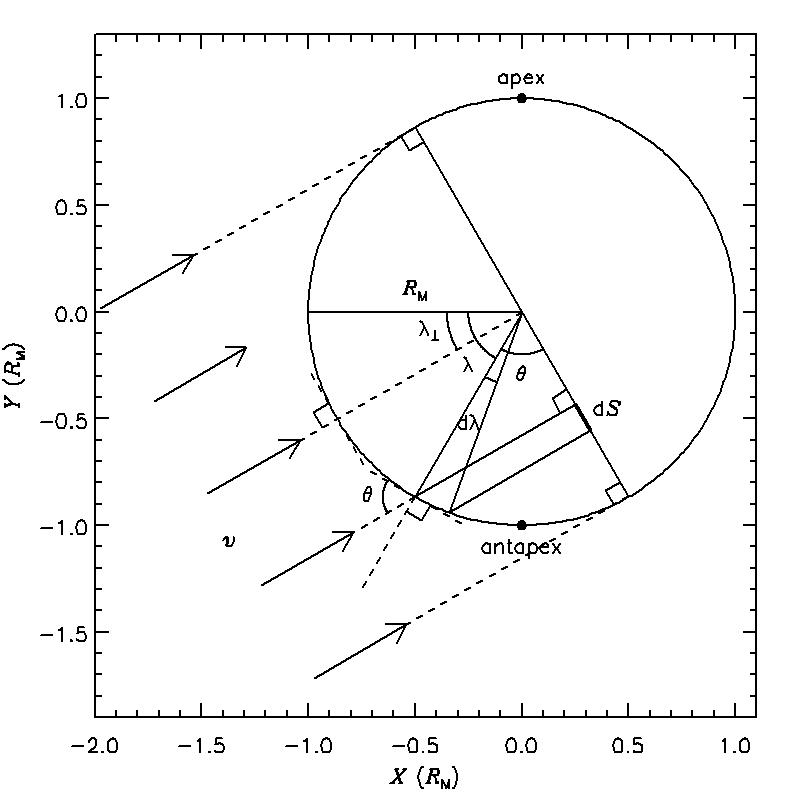}}
    \caption{Variables involved in integrating the cratering distribution. In the rest frame of the Moon, the common velocity of impactors is $\vec{v}$ (direction of the arrows). The hemisphere bounded by two positions where $\vec{v}$ (arrows on two sides) is parallel to the local horizons is the bombarded hemisphere, whose center, where $\vec{v}$ (arrow in the middle) is perpendicular to the lunar surface, is the normal impact point. Denotations are explained in the text.}
    \label{fig-ImpGeo2}
\end{figure}

When the encounter between one impactor orbit and the Earth orbit occurs, as seen in the Earth-Moon system, the impactors in the common orbit are treated as particles approaching along the parallel straight lines in the direction of their common encounter velocity (Fig. \ref{fig-ImpGeo}a).
They are assumed to be uniformly distributed in space, and to be numerous enough to cover the Moon's orbit, that is, the gravitational cross section of the Moon is always maximized (lunar diameter in the planar model).

In the geocentric frame with the x-axis pointing to the lunar perigee, the direction of $\vec{v}_{\rm{enc}}$ can be random. Since we average over the lunar period, it can be assumed to be in the positive y-axis direction without loss of generality. Figure \ref{fig-ImpGeo}a shows that the angle between the encounter velocity $\vec{v}_{\rm{enc}}$ and the lunar orbital velocity $\vec{v}_{\rm{M}}$ is the Moon's true anomaly $f_{\rm{M}}$, which varies uniformly at the rate of the mean motion $n_{\rm{M}}$. The impact velocity is the relative velocity between the impactors and the Moon:
\begin{align}
  \vec{v}_{\rm{enc}} &= (0, v_{\rm{enc}}), \\
  \vec{v}_{\rm{M}}   &= v_{\rm{M}} (-\sin f_{\rm{M}}, \cos f_{\rm{M}}), \\
  \vec{v} &= (v_{\rm{M}}\sin f_{\rm{M}}, v_{\rm{enc}}-v_{\rm{M}}\cos f_{\rm{M}}).
\end{align}
Its magnitude, that is, the impact speed, is
\begin{equation}    \label{eq-v_imp}
    v = \sqrt{v_{\rm{enc}}^2 + v_{\rm{M}}^2 -2 v_{\rm{enc}} v_{\rm{M}} \cos f_{\rm{M}}}
\end{equation}
At any instant, the impact can only occur on one hemisphere, and the impactors in this hemisphere have an equal $v$ since they share the same $v_{\rm{enc}}$. A normal impact point is the center of this bombarded hemisphere, where the incidence angle and thus the normal impact speed are both largest at this instant. As the impactors are assumed to be uniformly distributed, this is also where the impact flux is highest.

Given the encounter speed $v_{\rm{enc}}$ and lunar orbital speed $v_{\rm{M}}$, as long as $v_{\rm{enc}} > v_{\rm{M}}$, the impact speed $v$ is always minimized when the Moon is at the perigee ($f_{\rm{M}} = 0 \degr$) and maximized at the apogee ($f_{\rm{M}} = 180 \degr$), where the antapex and the apex become the normal impact points. Figure \ref{fig-ImpGeo} shows that the normal impact point constantly moves westward on the lunar surface at a varying rate, and that lower $v_{\rm{enc}}$ brings not only the lower $v$, but also a longer time spent on the leading side for the normal impact point, which implies that the lower the encounter speed, the stronger the leading/trailing asymmetry. That $v_{\rm{enc}} > v_{\rm{M}}$ is taken for granted hereafter in this work. This holds for the MBAs, whose minimum $v_{\rm{enc}}$ is 7.9 km~s$^{-1}$ ($a_{\rm{p}} = 2.5$ AU and $q_{\rm{p}} = 1$ AU), which is nearly equal to the maximum $v_{\rm{M}}$ when $a_{\rm{M}}$ = 1 $R_{\oplus}$. It is also valid for those NEOs with $a_{\rm{p}} > 1.2$ AU, whose minimum encounter speed $v_{\rm{enc}}$ = 2.4 km~s$^{-1}$ ($a_{\rm{p}} = 1.2$ AU and $q_{\rm{p}} = 1$ AU) approximates to $v_{\rm{M}}$ for $a_{\rm{M}} = 11 R_{\oplus}$, while $a_{\rm{M}}$ during the dominant epoch of the NEOs is at least about 40 $R_{\oplus}$.


Now we establish a rest frame of the Moon that is to be used in the integration. As shown in Fig. \ref{fig-ImpGeo2}, the positive y-axis points to the apex and the minus x-axis points to Earth. In this frame,
\begin{align}
  \vec{v} &= (v_x, v_y), \\
  v_x &= v_{\rm{enc}} \sin f_{\rm{M}}, \label{eq-v_x}  \\
  v_y &= v_{\rm{enc}} \cos f_{\rm{M}} - v_{\rm{M}} \label{eq-v_y},
\end{align}
while its magnitude $v$ keeps its expression of Eq. \ref{eq-v_imp}. We introduce some variables: $\lambda$ is the geometric longitude ranging from $-180\degr$ to $+180\degr$, measured eastward from the center of near side; $\theta$ is the incidence angle ranging from $0\degr$ to $90\degr$, the angle between $\vec{v}$ and the local horizon; $S$ is the cross section. A unit area (length in planar model) at longitude $\lambda$ on the bombarded hemisphere is
\begin{equation}
    {\rm{d}} l = R_{\rm{M}} {\rm{d}}\lambda
\end{equation}
and its cross section is
\begin{equation}
    {\rm{d}}S = R_{\rm{M}} \sin\theta {\rm{d}}\lambda.
\end{equation}
Denoting with $\vec{e} = (-\cos\lambda, -\sin\lambda)$ the normal vector, the incidence angle on the unit area can be derived with
\begin{equation}
    \sin\theta = \frac{\vec{e} \cdot \vec{v}}{v} = \frac{v_x \cos\lambda + v_y \sin\lambda}{v},
\end{equation}
and the normal speed $v_\bot$, the normal component of $\vec{v}$ is
\begin{equation}    \label{eq-vv}
    v_\bot = v \sin\theta = v_x \cos\lambda + v_y \sin\lambda.
\end{equation}
Denoting with $\rho$ the uniform spatial density of the impactors, the impact flux, the number of impactors received per unit time by the unit area is
\begin{equation}    \label{eq-dF}
    {\rm{d}}F = \rho v {\rm{d}}S = \rho R_{\rm{M}} v_\bot {\rm{d}}\lambda.
\end{equation}

At any instant when the lunar true anomaly is $f_{\rm{M}}$, the longitude of the normal impact point is defined as $\lambda_\bot$, where its normal vector $\vec{e}_\bot$ is parallel to $\vec{v}$, so that
\begin{align}
    &\sin\lambda_\bot = \frac{v_y(f_{\rm{M}})}{v(f_{\rm{M}})},   &\cos\lambda_\bot = \frac{v_x(f_{\rm{M}})}{v(f_{\rm{M}})}.
\end{align}
The bombarded hemisphere is then bounded by the longitude interval $[\lambda_{\rm{l}}, \lambda_{\rm{u}}]$, where the lower and upper limits $\lambda_{\rm{l,u}} = \lambda_\bot \mp 90\degr$, and
\begin{align}   \label{eq-lambda_lu}
    &\sin\lambda_{\rm{l,u}} = \mp\frac{v_x(f_{\rm{M}})}{v(f_{\rm{M}})},   &\cos\lambda_{\rm{l,u}} = \pm\frac{v_y(f_{\rm{M}})}{v(f_{\rm{M}})}.
\end{align}
As the bombarded hemisphere moves westward along the lunar equator, the moment a certain position $\lambda$ enters this hemisphere and the moment it leaves are when $\lambda$ becomes the lower and upper limits $\lambda_{\rm{l,u}}$, respectively, and when $\sin\theta = 0$ there. Therefore, the time interval while a certain position $\lambda$ is in the bombarded hemisphere, characterized by $[f_{\rm{l}}, f_{\rm{u}}]$, can be derived with
\begin{align}
    &\sin\lambda = \mp\frac{v_x(f_{\rm{l,u}})}{v(f_{\rm{l,u}})},   &\cos\lambda = \pm\frac{v_y(f_{\rm{l,u}})}{v(f_{\rm{l,u}})}.
\end{align}
It turns out
\begin{align}
  f_{\rm{l}} &= \arcsin (\frac{v_{\rm{M}}}{v_{\rm{enc}}}\sin\lambda) - \lambda, \\
  f_{\rm{u}} &= \pi - \arcsin (\frac{v_{\rm{M}}}{v_{\rm{enc}}}\sin\lambda) - \lambda.
\end{align}

\subsection{Exact formulations}   \label{subsec-exa}

\begin{figure*}
    \centering
    \includegraphics[width=17cm]{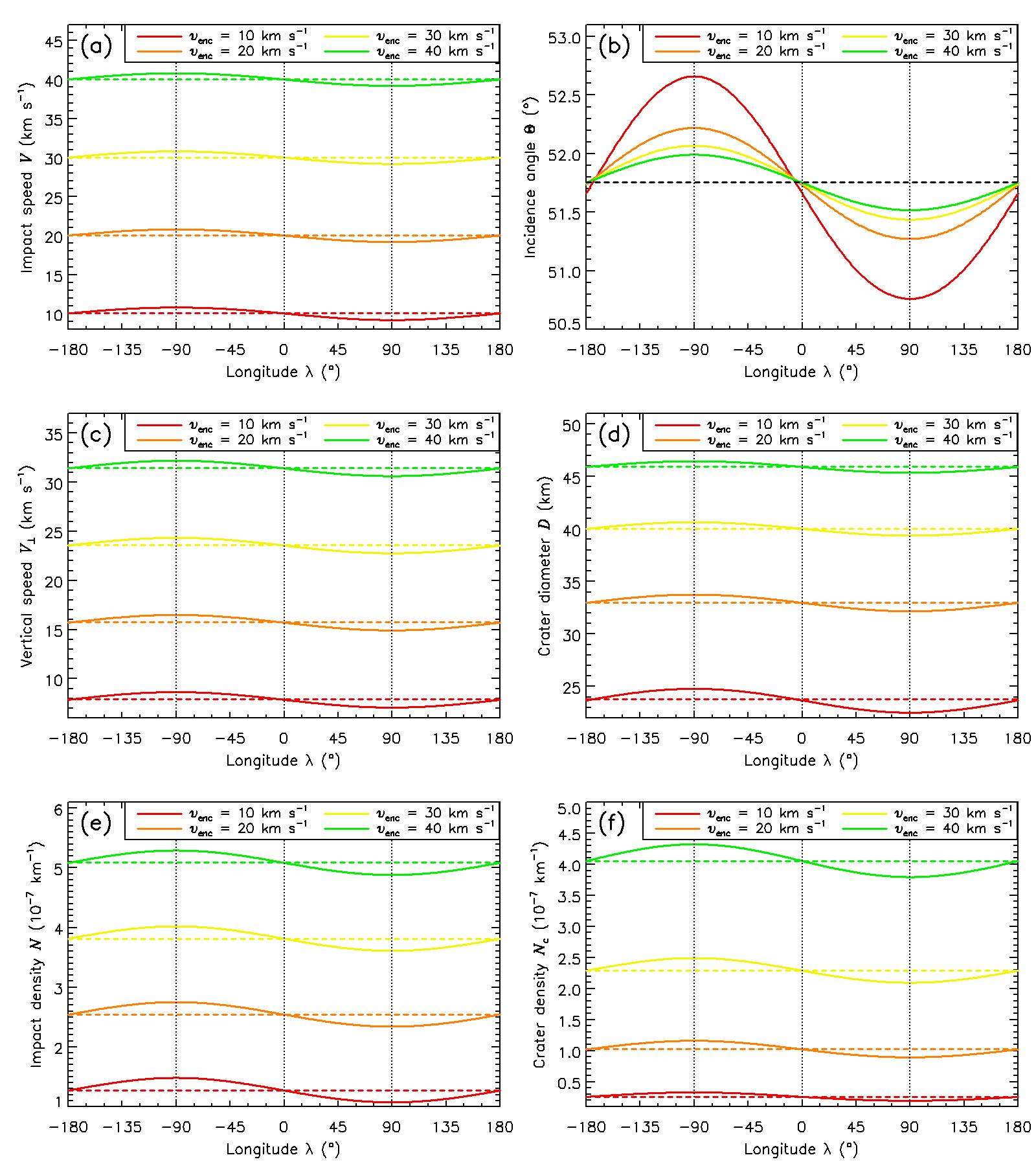}
    \caption{Impact speed (a), incidence angle (b), normal speed (c), crater diameter (d), impact density (e), and crater density (f) as functions of longitude. Their variations with longitude (solid curves in red, orange, yellow, and green for $v_{\rm{enc}}$ = 10, 20, 30, and 40 km~s$^{-1}$) and their global averages (dashed lines in the same color for the same $v_{\rm{enc}}$, except for those in panel b, which are all black to represent the invariant value) are calculated with $v_{\rm{M}} = 1.0$ km~s$^{-1}$, $d_{\rm{min}}$ = 0.5 km, $\alpha_{\rm{p}}$ = 1.75, $\rho = 3.4 \times 10^{-25}$ km$^{-2}$, $t = 3.7$ Gyr, and $d_{\rm{c}}$ = 25 km.}
    \label{fig-lam}
\end{figure*}

Here with $v_{\rm{enc}}$ and $v_{\rm{M}}$ given, the spatial distributions of impact density, impact speed, incidence angle, normal speed, crater diameter, and crater density are shown after integration. We note that the denotations of integrated variables are written uppercase to distinguish them from the instantaneous ones with lowercase letters.

The impact density $N$ is the number of impacts on a certain region divided by its area. For the unit area on longitude $\lambda$ illustrated in Fig. \ref{fig-ImpGeo2}, its impact density during a unit time is
\begin{equation}
    {\rm{d}} N = \frac{F {\rm{d}}t}{{\rm{d}}l} = \frac{\rho}{n_{\rm{M}}} (v_x \cos\lambda + v_y \sin\lambda) {\rm{d}}f_{\rm{M}}.
\end{equation}
Integration of ${\rm{d}} N$ over the interval $[f_{\rm{l}}, f_{\rm{u}}]$ leads to
\begin{equation}
    N (\lambda) = \frac{2 \rho}{n_{\rm{M}}} [\sqrt{v_{\rm{enc}}^2-v_{\rm{M}}^2\sin^2\lambda} - v_{\rm{M}} \sin\lambda \arccos(\frac{v_{\rm{M}}}{v_{\rm{enc}}} \sin\lambda)].
\end{equation}
To simplify the expression, we define $\sigma \in (0, \pi)$ by
\begin{equation}
    \sigma = \arccos(\frac{v_{\rm{M}}}{v_{\rm{enc}}} \sin\lambda),
\end{equation}
which leads to
\begin{eqnarray}
  \sin\sigma =& |\cos(f_{\rm{l,u}} + \lambda)| &= \sqrt{1 - (\frac{v_{\rm{M}}}{v_{\rm{enc}}} \sin\lambda)^2}, \\
  \cos\sigma =& \sin(f_{\rm{l,u}} + \lambda) &= \frac{v_{\rm{M}}}{v_{\rm{enc}}} \sin\lambda.
\end{eqnarray}
Therefore,
\begin{equation}    \label{eq-N lambda per}
    N (\lambda) = \frac{2 \rho v_{\rm{enc}}}{n_{\rm{M}}} (\sin\sigma - \sigma \cos\sigma).
\end{equation}

The impact speed $V$, incidence angle $\Theta$, and normal speed $V_\bot$, as functions of longitude, are averages of $v$, $\theta$, and $v_\bot$ over one lunar period, that is, $V = (\int_{f_{\rm{l}}}^{f_{\rm{u}}} v {\rm{d}}N) /N$, $\sin\Theta = (\int_{f_{\rm{l}}}^{f_{\rm{u}}} \sin\theta {\rm{d}}N) /N$, and $V_\bot = (\int_{f_{\rm{l}}}^{f_{\rm{u}}} v_\bot {\rm{d}}N) /N$. The integration leads to
\begin{align}
  V(\lambda) &= I_1/I_0,   \label{eq-V lambda} \\
  \sin\Theta(\lambda) &=  I_2/I_0,   \label{eq-Theta lambda} \\
  V_\bot(\lambda) &= I_3/I_0  \label{eq-VV lambda},
\end{align}
where
\begin{align}
  I_0 &= 2 v_{\rm{enc}} (\sin\sigma - \sigma \cos\sigma),  \label{eq-I0}    \\
\begin{split}
  I_1 &= \frac{1}{3v_{\rm{M}}}
        \{[(v_{\rm{enc}}^2+v_{\rm{M}}^2) \cos\lambda + 2 v_{\rm{M}}v_{\rm{enc}} \sin\sigma] I_+ \\
      &- [(v_{\rm{enc}}^2+v_{\rm{M}}^2)\cos\lambda - 2 v_{\rm{M}}v_{\rm{enc}} \sin\sigma] I_- \\
      &+ (v_{\rm{enc}} + v_{\rm{M}})(v_{\rm{enc}}^2 + 7v_{\rm{M}}^2) (\sin\lambda) \Delta E \\
      &- (v_{\rm{enc}} + v_{\rm{M}}) (v_{\rm{enc}} - v_{\rm{M}})^2 (\sin\lambda) \Delta F \},
\end{split} \\
\begin{split}
  I_2 &=  \frac{1}{3v_{\rm{M}}^2}
        \{[(2v_{\rm{enc}}^2 - 3v_{\rm{M}}^2)\cos\lambda - v_{\rm{M}} v_{\rm{enc}} \sin\sigma] (\sin\lambda) I_+ \\
    & - [(2v_{\rm{enc}}^2 - 3v_{\rm{M}}^2)\cos\lambda + v_{\rm{M}} v_{\rm{enc}} \sin\sigma] (\sin\lambda) I_-  \\
    & - (v_{\rm{enc}} + v_{\rm{M}}) [(v_{\rm{enc}}^2-2v_{\rm{M}}^2)\cos(2\lambda) + 3v_{\rm{M}}^2] \Delta E \\
    & + (v_{\rm{enc}} + v_{\rm{M}}) (v_{\rm{enc}} - v_{\rm{M}})^2 \cos(2\lambda) \Delta F \},
\end{split} \\
  I_3 &= v_{\rm{enc}}^2 [(1 + 2 \cos^2\sigma) \sigma - 3 \sin\sigma \cos\sigma],    \label{eq-I3}
\end{align}
\begin{align}
  I_{\pm} &= \sqrt{v_{\rm{enc}}^2 + v_{\rm{M}}^2 \cos(2\lambda) \pm 2 v_{\rm{M}} v_{\rm{enc}} \cos\lambda \sin\sigma},
\end{align}
\begin{align}
  \Delta E &= E(\frac{\lambda-\sigma}{2} + \frac{\pi}{4} | k^2) - E(\frac{\lambda + \sigma}{2} + \frac{\pi}{4} - \delta\pi | k^2) - 2\delta E(k^2), \\
  \Delta F &= F(\frac{\lambda-\sigma}{2} + \frac{\pi}{4} | k^2) - F(\frac{\lambda + \sigma}{2} + \frac{\pi}{4} - \delta\pi | k^2) - 2\delta F(k^2), \\
  k^2 &= \frac{4 v_{\rm{M}} v_{\rm{enc}}}{(v_{\rm{enc}} + v_{\rm{M}})^2},  \label{eq-k^2}  \\
  \delta &=
  \begin{cases}
    0,     & (\lambda < 0)         \\
    1.     & (\lambda > 0)
  \end{cases}
\end{align}
Functions $F(\phi|k^2)$ and $F(k^2)$ are the incomplete and complete elliptic integrals of the first kind; $E(\phi|k^2)$ and $E(k^2)$ are those of the second kind.

The crater size generated by an impact depends on the impact speed, incidence angle, the projectile size, the target surface gravity, the densities of the two objects, and so on. Assuming the cratering efficiency of an oblique impact depends on the normal speed, the crater scaling law allows converting between crater diameter $d_{\rm{c}}$ and projectile diameter $d_{\rm{p}}$ with
\begin{align}
  d_{\rm{p}} (d_{\rm{c}}, v_\bot) &
  = c_{\rm{p}} d_{\rm{c}}^{\gamma'_{\rm{p}}} v_\bot^{-\gamma_{\rm{p}}}, \label{eq-d_p}  \\
  d_{\rm{c}}(d_{\rm{p}}, v_\bot) &
  = c_{\rm{c}} d_{\rm{p}}^{\gamma'_{\rm{c}}} v_\bot^{\gamma_{\rm{c}}}.  \label{eq-d_c}
\end{align}
The constants $c_{\rm{p,c}}$, $\gamma_{\rm{p,c}}$, and $\gamma'_{\rm{p,c}}$ given by different studies vary, but their values are not important to the analytical derivation here. Still, they are needed for illustration and post-processing of numerical simulations, and we therefore adopt the forms given by \citet{LeFeuvre2011} for a non-porous regime (working for $d_{\rm{c}} > 20$ km) with the Moon as the target:
\begin{align}
  c_{\rm{p}} &= [\frac{0.98}{1.56K} d_*^{0.079} g^{\nu_1} (\rho_{\rm{t}} / \rho_{\rm{p}})^{\nu_2}]^{\frac{1}{1-\nu_1}}, \label{eq-c_p} \\
  \gamma_{\rm{p}} &= \frac{2 \nu_1}{1 - \nu_1}, \label{eq-gamma_p} \\
  \gamma'_{\rm{p}} &= \frac{0.921}{1 - \nu_1}  \label{eq-gamma'_p}, \\
  c_{\rm{c}} &= [\frac{1.56 K g^{-\nu_1} (\rho_{\rm{t}} / \rho_{\rm{p}})^{-\nu_2}}{0.98 d_*^{0.079}}]^{\frac{1}{0.921}}, \label{eq-c_c}  \\
  \gamma_{\rm{c}} &= \frac{2\nu_1}{0.921}, \label{eq-gamma_c} \\
  \gamma'_{\rm{c}} &= \frac{1-\nu_1}{0.921}  \label{eq-gamma'_c},
\end{align}
where the non-porous scaling parameters are $K = 1.17$, $\nu_1 = 0.22$, and $\nu_2 = 0.31$, the lunar surface gravity is $g = 1.6$ m~s$^{-2}$, the target density is $\rho_{\rm{t}} = 2.8$ g~cm$^{-3}$, the projectile density is $\rho_{\rm{p}} = 3$ g~cm$^{-3}$, and the transition diameter is $d_* = 8.5$ km for the Moon. Again we point out that the following derivation is based on Eqs. \ref{eq-d_p} and \ref{eq-d_c}, but not on Eqs. \ref{eq-c_p}--\ref{eq-gamma'_c}.

The cumulative size distribution of an impactor population can be commonly assumed as $N_{\rm{p}} \propto d_{\rm{p}}^{-\alpha_{\rm{p}}}$, independent of its orbital distribution. Given a minimum of projectile size $d_{\rm{min}}$, a normalized size distribution is
\begin{equation}    \label{eq-N_p}
    \bar{N}_{\rm{p}}(>d_{\rm{p}}) = (\frac{d_{\rm{p}}}{d_{\rm{min}}})^{-\alpha_{\rm{p}}}, \quad (d_{\rm{p}} \ge d_{\rm{min}}).
\end{equation}
It can be interpreted as the probability of an arbitrary impactor to be larger than $d_{\rm{p}}$. The mean size of the projectiles is
\begin{equation}    \label{eq-d_p mean}
    \bar{d}_{\rm{p}} = \int_{+\infty}^{d_{\rm{min}}} d_{\rm{p}} {\rm{d}}\bar{N}_{\rm{p}} = \frac{\alpha_{\rm{p}}}{\alpha_{\rm{p}}-1}d_{\rm{min}},
\end{equation}
where the slope $\alpha_{\rm{p}} > 1$ (otherwise, the mean projectile size would be infinite). This equation should hold everywhere on the Moon since the size distribution is independent of orbital distribution, and gravitations on impactors are ignored. Therefore, the spatial distribution of the periodic averaged crater diameter is $D = [\int_{f_{\rm{l}}}^{f_{\rm{u}}} d_{\rm{c}}(\bar{d}_{\rm{p}}, v_\bot) {\rm{d}}N] /N$, which is substituted with $d_{\rm{c}}(\bar{d}_{\rm{p}}, V_\bot(\lambda))$ to avoid difficult analytical integration. The deviation of the substitute is no more than 1.5\% for the ratio $v_{\rm{M}}/v_{\rm{enc}}$ as high as 0.25 (the case $a_{\rm{M}}$ = 10 $R_{\oplus}$ and $v_{\rm{enc}} = 10$ km~s$^{-1}$) and even lower for a lower speed ratio. Thus we derive
\begin{equation}    \label{eq-D lambda}
    D(\lambda) = c_{\rm{c}} \bar{d}_{\rm{p}}^{\gamma'_{\rm{c}}} [V_\bot (\lambda)]^{\gamma_{\rm{c}}}.
\end{equation}

The crater density $N_{\rm{c}}(>d_{\rm{c}})$ is the number of craters with diameters larger than $d_{\rm{c}}$ per unit area. It is relevant to the age determination in the cratering chronology method. It depends on the projectile diameter $d_{\rm{p}}$, the projectile size-frequency distribution $\bar{N}_{\rm{p}}$, and the impact density $N$. On the assumption that every impact leaves one and only one crater, meaning that saturation, erosion, and secondary craters are all ignored,
\begin{equation}
{\rm{d}}N_{\rm{c}}(>d_{\rm{c}}) = \bar{N}_{\rm{p}}(>d_{\rm{p}}(d_{\rm{c}}, v_\bot)) {\rm{d}}N.
\end{equation}
In every unit time ${\rm{d}}f_{\rm{M}} \in [f_{\rm{l}}, f_{\rm{u}}]$, the factor ${\rm{d}}N$ gives the total density of craters on the longitude $\lambda$; function $d_{\rm{p}}$ gives the projectile size required to form a crater as large as $d_{\rm{c}}$ there; function $\bar{N}_{\rm{p}}$ gives the fraction of impactors larger than $d_{\rm{p}}$, which is equivalent to the fraction of craters larger than $d_{\rm{c}}$; and finally the product of $\bar{N}_{\rm{p}}$ and ${\rm{d}}N$ determines the required part of ${\rm{d}}N$. The substitute for the integral $\int_{f_{\rm{l}}}^{f_{\rm{u}}} \bar{N}_{\rm{p}}(>d_{\rm{p}}(d_{\rm{c}}, v_\bot)) {\rm{d}}N$ is $\bar{N}_{\rm{p}}(> d_{\rm{p}}(d_{\rm{c}}, V_\bot(\lambda))) N (\lambda)$, whose deviation is no more than 0.07\% for a ratio $v_{\rm{M}}/v_{\rm{enc}}$ as high as 0.25 and even smaller for a lower speed ratio. Assuming $d_{\rm{c}}$ is large enough to ensure $d_{\rm{p}}(d_{\rm{c}}, V_\bot(\lambda)) \ge d_{\rm{min}}$ holds everywhere on the lunar surface, then
\begin{equation}    \label{eq-N_c lambda}
  N_{\rm{c}}(>d_{\rm{c}}, \lambda)
    = (\frac{d_{\rm{min}}}{c_{\rm{p}} d_{\rm{c}}^{\gamma'_{\rm{p}}}})^{\alpha_{\rm{p}}} [V_\bot(\lambda)]^{\gamma_{\rm{p}} \alpha_{\rm{p}}} N(\lambda).
\end{equation}
We note that $N_{\rm{c}}(>d_{\rm{c}}, \lambda)$ describes not only the spatial distribution of the crater density, but also the craters' size distribution. Equation \ref{eq-N_c lambda} proves $N_{\rm{c}} \propto d_{\rm{c}}^{- \gamma'_{\rm{p}} \alpha_{\rm{p}}}$ on any lunar longitude, and thus the slopes of the size distributions of impactors and craters are related by
\begin{equation}    \label{eq-slope p2c}
    \alpha_{\rm{c}} = \gamma'_{\rm{p}} \alpha_{\rm{p}}.
\end{equation}

Although the integration is made over one lunar period, all of the distributions but $N(\lambda)$ and $N_{\rm{c}}(>d_{\rm{c}}, \lambda)$ have no dependence on time, so that they are applicable to any time interval $t$ (multiple of lunar period, theoretically) as long as $v_{\rm{M}}$ and $v_{\rm{enc}}$ are constant. We rewrite $N(\lambda)$ after multiplying it by $n_{\rm{M}}t/(2\pi)$:
\begin{equation}    \label{eq-N lambda}
  N(\lambda) = \frac{\rho t v_{\rm{enc}}}{\pi} (\sin\sigma - \sigma \cos\sigma).
\end{equation}
This form describes the impact density after a bombardment duration $t$, but its relative variation does not differ at all from the one-periodic form, which is one of the reasons we suggest using the asymmetry amplitude to measure the relative variation (Sect. \ref{subsec-appr}). Hereafter, this new expression of $N(\lambda)$ takes the place of the previous one, while $N_{\rm{c}}(>d_{\rm{c}}, \lambda)$ (Eq. \ref{eq-N_c lambda}) is still valid.

All of the absolute spatial distributions are shown in Fig. \ref{fig-lam} with solid curves. We caution that on each unit area, there must be impacts with all the (instantaneous) incidence angles $\theta \in$ [0$\degr$, 90$\degr$] theoretically, and thus all the normal speeds $v_\bot \in$ [0, $v$] and crater diameters $d_{\rm{c}} \in$ [0, $d_{\rm{c}}(\bar{d}_{\rm{p}}, v)$], nevertheless, their periodic averages are what the figure shows. The most significant common feature is that the curves all peak at the apex ($\lambda = - 90\degr$) and are lowest at the antapex ($\lambda = + 90\degr$), regardless of $v_{\rm{enc}}$. The leading/trailing asymmetry is unambiguously detected in distributions of all the investigated variables, which is to be confirmed again through approximate expressions (Sect. \ref{subsec-appr}). Additionally, it is apparent that higher $v_{\rm{enc}}$ leads to an upward shift of all the curves, expect for $\Theta (\lambda)$. An increase in $v_{\rm{enc}}$ diminishes and enlarges the absolute amplitudes of $D$ and $N_{\rm{c}}$, while it almost does not influence those of $N$, $V$, and $V_\bot$. With $v_{\rm{M}}$ fixed, the absolute amplitude of $\Theta$ that only depends on the speed ratio $v_{\rm{M}}/v_{\rm{enc}}$ also seems to decrease as $v_{\rm{enc}}$ increases. We provide the explanation in Sect. \ref{subsec-appr}.

\subsection{Near/far symmetry}   \label{subsec-near/far}

\begin{figure*}
    \centering
    \includegraphics[width=17cm]{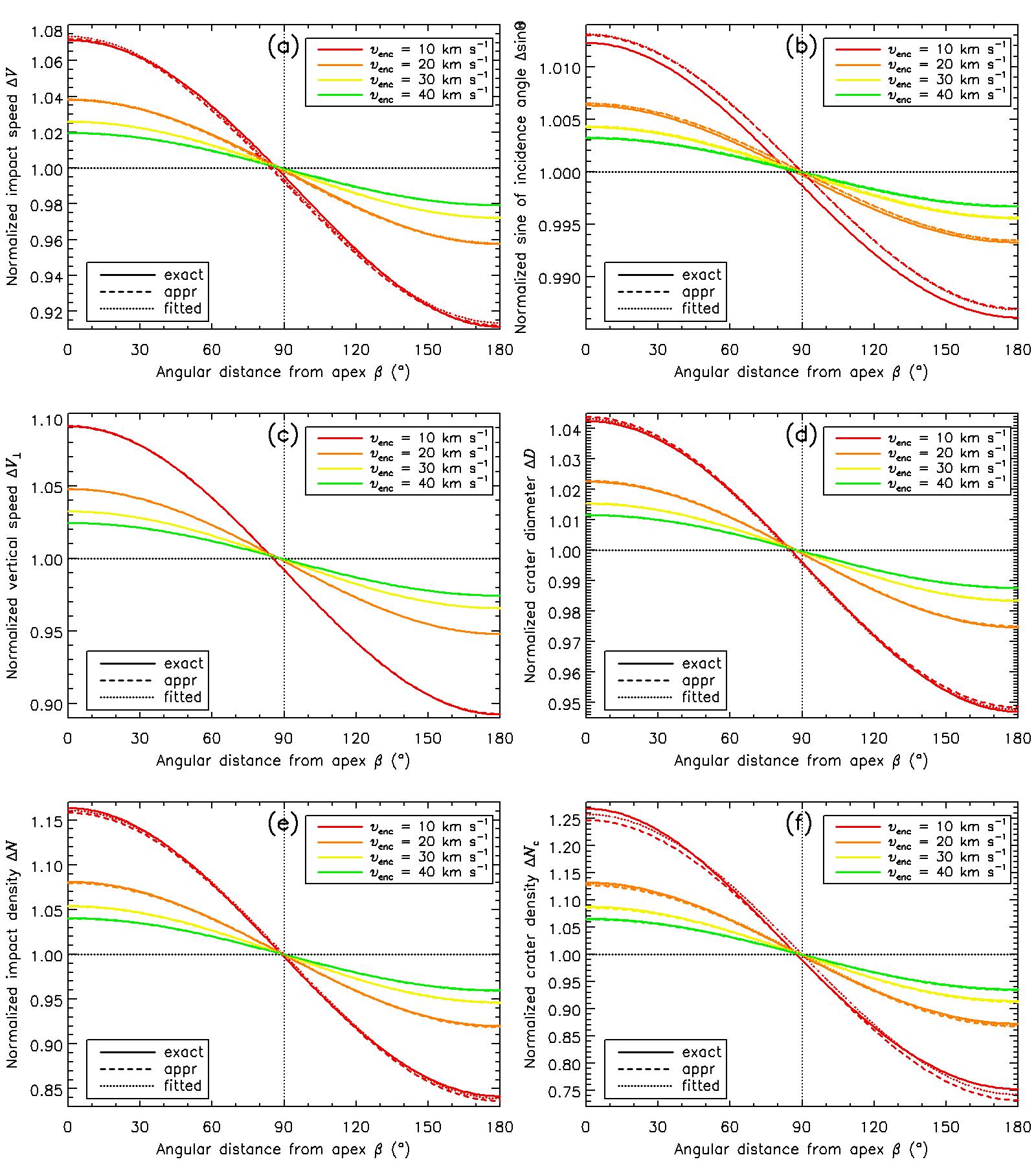}
    \caption{Normalized impact speed (a), incidence angle (b), normal speed (c), crater diameter (d), impact density (e), and crater density (f) as functions of $\beta$. The exact variations with $\beta$ (solid curves in red, orange, yellow, and green for $v_{\rm{enc}}$ = 10, 20, 30, and 40 km~s$^{-1}$) are calculated using exact formulations with $v_{\rm{M}} = 1.0$ km~s$^{-1}$, $d_{\rm{min}}$ = 0.5 km, $\alpha_{\rm{p}}$ = 1.75, $\rho = 3.4 \times 10^{-25}$ km$^{-2}$, $t = 3.7$ Gyr, and $d_{\rm{c}}$ = 25 km. The approximate variations (dashed curves in the same color as the solid curves for the same $v_{\rm{enc}}$) are calculated with the same parameters but using the approximate series of formulations. The fit variations (dotted curves in the same color as the solid curves for the same $v_{\rm{enc}}$) are best fits of exact variations to the approximate formulations. All the variations are in terms of the relevant exact global averages (horizontal dotted lines).}
    \label{fig-beta}
\end{figure*}

Because $N(\lambda) = N(\pm180\degr - \lambda)$, $N(\lambda)$ is symmetric about $\lambda = \pm 90\degr$, meaning that the distribution of the impact density on the lunar surface is symmetric about the line connecting the apex and the antapex. The same is easily found in terms of $V$, $\Theta$, $V_\bot$, $D$, and $N_{\rm{c}}$. Therefore, the Moon's near and far sides are exact mirror images of each other, with no sign of a near/far asymmetry.

This conclusion is based on the assumption that the Earth's gravitation on the impactors and its volume are not considered, meaning that Earth is treated as if it were transparent to the impactors. In reality, it may block impactors like a shield, preventing them from impacting the Moon, or gravitationally focusing them like a lens, increasing the flux that the near side receives. \citet{Bandermann1973} confirmed the former effect for the condition that $a_{\rm{M}} < 25 R_{\oplus}$;
\citet{Gallant2009} claimed that there was very little asymmetry for $a_{\rm{M}}$ in the range 10--50 $R_{\oplus}$; \citet{LeFeuvre2011} found the asymmetry negligible with about 0.1\% more craters of the near side than the far side, in contrast with \citet{LeFeuvre2005}, who claimed a factor of four enhancement on the near side using impactors coplanar with Earth. Our deduction shows that without the Earth's effect, the near/far asymmetry cannot exist, while the leading/trailing asymmetry is the inherent and natural consequence of cratering.

We therefore use $\beta$, the angular distance from the apex, to describe the leading/trailing asymmetry and not the longitude $\lambda$. It ranges between 0$\degr$ and 180$\degr$, where the apex and antapex are, respectively. For the near side, where $-90\degr \le \lambda \le +90\degr$,
\begin{equation*}
    \lambda = \beta - 90\degr.
\end{equation*}
Substituting $\lambda$ with $\beta$, the integrated variables as functions of $\beta$ are
\begin{align}
  N (\beta) &= \frac{\rho t v_{\rm{enc}}}{\pi} (\sin\sigma - \sigma \cos\sigma),  \label{eq-N beta}   \\
  V(\beta) &= I_1'/I_0',   \label{eq-V beta}   \\
  \sin\Theta(\beta) &= I_2'/I_0', \label{eq-Theta beta}    \\
  V_\bot(\beta) &= I_3'/I_0', \label{eq-VV beta}   \\
  D(\beta) &= c_{\rm{c}} \bar{d}_{\rm{p}}^{\gamma'_{\rm{c}}} [V_\bot (\beta)]^{\gamma_{\rm{c}}},  \label{eq-D beta}   \\
  N_{\rm{c}}(>d_{\rm{c}}, \beta) &= (\frac{d_{\rm{min}}}{c_{\rm{p}} d_{\rm{c}}^{\gamma'_{\rm{p}}}})^{\alpha_{\rm{p}}} [V_\bot(\beta)]^{\gamma_{\rm{p}} \alpha_{\rm{p}}} N(\beta)  \label{eq-N_c beta},
\end{align}
where
\begin{align}
  I_0' &= 2 (\sin\sigma - \sigma \cos\sigma),   \\
\begin{split}
  I_1' &= \frac{v_{\rm{enc}}}{3\eta}
        \{[(1+\eta^2) \sin\beta + 2 \eta \sin\sigma] I_+' \\
      &- [((1+\eta^2) \sin\beta - 2 \eta \sin\sigma] I_-' \\
      &- (1 + \eta)(1 + 7\eta^2) (\cos\beta) \Delta E \\
      &+ (1 + \eta) (1 - \eta)^2 (\cos\beta) \Delta F \},
\end{split} \\
\begin{split}
  I_2' &=  \frac{1}{3\eta^2}
        \{-[(2 - 3\eta^2)\sin\beta - \eta \sin\sigma] (\cos\beta) I_+' \\
    & + [(2 - 3\eta^2)\sin\beta + \eta \sin\sigma] (\cos\beta) I_-'  \\
    & + (1 + \eta) [(1-2\eta^2)\cos(2\beta) - 3\eta^2] \Delta E \\
    & - (1 + \eta) (1 - \eta)^2 \cos(2\beta) \Delta F \},
\end{split} \\
  I_3' &= v_{\rm{enc}} [(1 + 2 \cos^2\sigma) \sigma - 3 \sin\sigma \cos\sigma], \\
  I_{\pm}' &= \sqrt{1 - \eta^2 \cos(2\beta) \pm 2 \eta \sin\beta \sin\sigma},
\end{align}
\begin{align}
  \Delta E &= E(\frac{\beta-\sigma}{2} | k^2) - E(\frac{\beta + \sigma}{2} - \delta\pi | k^2) - 2\delta E(k^2), \\
  \Delta F &= F(\frac{\beta-\sigma}{2} | k^2) - F(\frac{\beta + \sigma}{2} - \delta\pi | k^2) - 2\delta F(k^2), \\
  k^2 &= \frac{4\eta}{(1 + \eta)^2}, \\
  \delta &=
  \begin{cases}
    0,     & (\beta < \frac{\pi}{2})         \\
    1,     & (\beta > \frac{\pi}{2})
  \end{cases}
\end{align}
\begin{align}
  \sigma &= \arccos(- \eta \cos\beta), \\
  \sin\sigma &= \sqrt{1 - (\eta \cos\beta)^2}, \\
  \cos\sigma &= - \eta \cos\beta,  \\
  \eta &= \frac{v_{\rm{M}}}{v_{\rm{enc}}}.
\end{align}
We note that the above equations are applicable to both the near and the far side.

Figure \ref{fig-beta} illustrates the relative spatial variations of $N$, $V$, $\Theta$, $V_\bot$, $D$, and $N_{\rm{c}}$ as functions of $\beta$ with solid curves. The relative variation is the absolute variation divided by its global average (to be derived in Sect. \ref{subsec-glo}). All variables decrease monotonically with increasing $\beta$, as we show in Fig. \ref{fig-lam}. Moreover, the greater the encounter speed $v_{\rm{enc}}$ (the lower the speed ratio $v_{\rm{M}}/v_{\rm{enc}}$ with given $v_{\rm{M}}$), the smaller the relative variation amplitudes (to be explained in Sect. \ref{subsec-appr}).

\subsection{Global averages}   \label{subsec-glo}

Here we derive the global averages of the impact density $N$, impact speed $V$, incidence angle $\Theta$, normal speed $V_\bot$, crater diameter $D$, and crater density $N_{\rm{c}}$ through integrating.
The impact number for a unit area on the longitude $\lambda$ during a unit time is
\begin{equation}
    {\rm{d}}^2 C = {\rm{d}}F {\rm{d}}t = \frac{\rho R_{\rm{M}}}{n_{\rm{M}}} (v_x \cos\lambda + v_y \sin\lambda) {\rm{d}}\lambda {\rm{d}}f_{\rm{M}},
\end{equation}
where $F$ is the impact flux (Eq. \ref{eq-dF}), $v_{x,y}$ are elements of $\vec{v}$ (Eqs. \ref{eq-v_x} and \ref{eq-v_y}), and $\rho$ is the uniform spatial density of impactors. The total number of impacts on the global lunar surface after one lunar period is the integral $\iint {\rm{d}}^2 C$ over the $\lambda$ interval $[\lambda_{\rm{l}}, \lambda_{\rm{u}}]$ and $f_{\rm{M}}$ interval $[0\degr, 360\degr]$ in turn, where $\lambda_{\rm{l,u}} (f_{\rm{M}})$ are boundaries of the instantaneous bombarded hemisphere (Eq. \ref{eq-lambda_lu}). Thus,
\begin{equation}
    C = \frac{8 \rho R_{\rm{M}}}{n_{\rm{M}}} (v_{\rm{enc}}+v_{\rm{M}}) E(k^2).
\end{equation}
The global average of $N$ for the bombardment duration $t$ (multiple of lunar period) is the total impact number for one period, $C$, divided by the global area $l_{\rm{gl}} = 2 \pi R_{\rm{M}}$ (perimeter for the planar model) and multiplied by $n_{\rm{M}}t/(2\pi) $:
\begin{equation}    \label{eq-N glo}
    \bar{N} = \frac{2 \rho t}{\pi^2} (v_{\rm{enc}}+v_{\rm{M}}) E(k^2).
\end{equation}

Apparently, the global averages of $V$, $\Theta$, and $V_\bot$ are the integrals $\iint v {\rm{d}}^2C$, $\iint \sin\theta {\rm{d}}^2C$, and $\iint v_\bot {\rm{d}}^2C$ over the above $\lambda$ and $f_{\rm{M}}$ intervals divided by $C$, respectively. The integration results in
\begin{align}
  \bar{V} &= \frac{\pi}{2} \frac{v_{\rm{enc}}^2+v_{\rm{M}}^2}{(v_{\rm{enc}}+v_{\rm{M}}) E(k^2)},  \label{eq-V glo}  \\
  \sin\bar{\Theta} &= \frac{\pi}{4},  \label{eq-Theta glo}  \\
  \bar{V}_\bot &= \frac{\pi^2}{8} \frac{v_{\rm{enc}}^2+v_{\rm{M}}^2}{(v_{\rm{enc}}+v_{\rm{M}}) E(k^2)} \label{eq-VV glo},
\end{align}
where $\bar{\Theta}$ is defined by $\sin\bar{\Theta} = \overline{\sin\theta}$, and it is found $\bar{V}_\bot = \bar{V}\sin\bar{\Theta}$.
The global averages of $D$ and $N_{\rm{c}}$ are $[\iint d_{\rm{c}}(\bar{d}_{\rm{p}}, v_\bot) {\rm{d}}^2C]/C$ and $[\iint \bar{N}_{\rm{p}}(>d_{\rm{p}}(d_{\rm{c}}, v_\bot)) {\rm{d}}^2C]/l_{\rm{gl}}$, but to avoid analytical integration, we again substitute them by $d_{\rm{c}}(\bar{d}_{\rm{p}}, \bar{V}_\bot)$ and $\bar{N}_{\rm{p}}(> d_{\rm{p}}(d_{\rm{c}}, \bar{V}_\bot)) \bar{N}$, functions (Eqs. \ref{eq-d_c} and \ref{eq-d_p}) of $\bar{V}_\bot$ and $\bar{N}$. The substitutes are quite acceptable, for their deviations are $< 3\%$ and $< 1\%$, respectively, for a speed ratio $v_{\rm{M}}/v_{\rm{enc}}$ as high as 0.25 (the case $a_{\rm{M}}$ = 10 $R_{\oplus}$ and $v_{\rm{enc}}$ = 10 km~s$^{-1}$). A lower speed ratio leads to even smaller deviations. Thus, we derive
\begin{align}
  \bar{D} &=  c_{\rm{c}} \bar{d}_{\rm{p}}^{\gamma'_{\rm{c}}} \bar{V}_\bot^{\gamma_{\rm{c}}},  \label{eq-D glo} \\
  \bar{N}_{\rm{c}} &= (\frac{d_{\rm{min}}}{c_{\rm{p}} d_{\rm{c}}^{\gamma'_{\rm{p}}}})^{\alpha_{\rm{p}}} \bar{V}_\bot^{\gamma_{\rm{p}} \alpha_{\rm{p}}} \bar{N} \label{eq-N_c glo}.
\end{align}

The global averages are shown in Fig. \ref{fig-lam} with horizontal dashed lines. They all exhibit positive relations with $v_{\rm{enc}}$, except for $\bar{\Theta}$. Equation \ref{eq-Theta glo} explains the exception by determining the global average of $\sin\theta$ to be $\pi/4$ always, equivalent to $\bar{\Theta} = 51.8\degr$, regardless of both the encounter speed $v_{\rm{enc}}$ and the lunar orbital speed $v_{\rm{M}}$.

\subsection{Approximate formulations}   \label{subsec-appr}

We have formulated the spatial distributions $N(\beta)$, $V(\beta)$, $\Theta(\beta)$, $V_\bot(\beta)$, $D(\beta)$, and $N_{\rm{c}}(>d_{\rm{c}},\beta)$ (Eqs. \ref{eq-N beta}--\ref{eq-N_c beta}) and the global averages $\bar{N}$, $\bar{V}$, $\bar{V}_\bot$, $\bar{\Theta}$, $\bar{D}$, and $\bar{N}_{\rm{c}}$ (Eqs. \ref{eq-N glo}--\ref{eq-N_c glo}). Hereafter, we use $\Gamma$ to denote any one (or every one) of the variables $N$, $V$, $\Theta$, $V_\bot$, $D$, and $N_{\rm{c}}$, and use $\bar{\Gamma}$ to denote its global average. The normalized distribution is defined as the absolute distribution divided by the global average, which describes the relative variation
\begin{equation}
    \Delta\Gamma (\beta) = \Gamma(\beta) /\bar{\Gamma}.
\end{equation}
The above variables represent every aspect of cratering and their spatial distributions are what we call cratering distribution. Each of the formulations $\Gamma (\beta)$ and $\bar{\Gamma}$ except for the constant $\bar{\Theta}$ can be taken as the product of two factors involving the encounter speed $v_{\rm{enc}}$ and the speed ratio $\eta = v_{\rm{M}}/v_{\rm{enc}}$ respectively. Recalling the assumption $v_{\rm{M}}<v_{\rm{enc}}$, we can derive their series expansions around $\eta = 0$. To first order, the formulations are simplified to be
\begin{align}
  N(\beta) &= \frac{\rho t v_{\rm{enc}}}{\pi} (1 + \frac{\pi}{2} \frac{v_{\rm{M}}}{v_{\rm{enc}}} \cos\beta), \label{eq-N appr} \\
  V(\beta) &= v_{\rm{enc}} (1 + \frac{\pi}{4} \frac{v_{\rm{M}}}{v_{\rm{enc}}} \cos\beta), \label{eq-V appr}    \\
  \sin\Theta(\beta) &= \frac{\pi}{4} (1 + \frac{32-3\pi^2}{6\pi} \frac{v_{\rm{M}}}{v_{\rm{enc}}} \cos\beta),  \label{eq-Theta appr} \\
  V_\bot(\beta) &= \frac{\pi v_{\rm{enc}}}{4} (1 + \frac{16-\pi^2}{2\pi} \frac{v_{\rm{M}}}{v_{\rm{enc}}} \cos\beta), \label{eq-VV appr} \\
  D(\beta) &= c_{\rm{c}} \bar{d}_{\rm{p}}^{\gamma'_{\rm{c}}} (\frac{\pi v_{\rm{enc}}}{4})^{\gamma_{\rm{c}}} [1 + \frac{(16-\pi^2)\gamma_{\rm{c}}}{2\pi} \frac{v_{\rm{M}}}{v_{\rm{enc}}} \cos\beta], \label{eq-D appr} \\
  N_{\rm{c}}(\beta) &= \frac{\rho t v_{\rm{enc}}}{\pi} [\frac{d_{\rm{min}}}{c_{\rm{p}} d_{\rm{c}}^{\gamma'_{\rm{p}}}} (\frac{\pi v_{\rm{enc}}}{4})^{\gamma_{\rm{p}}}]^{\alpha_{\rm{p}}} [1 + \frac{\pi^2 + (16-\pi^2)\gamma_{\rm{p}}\alpha_{\rm{p}}}{2\pi} \frac{v_{\rm{M}}}{v_{\rm{enc}}} \cos\beta] \label{eq-N_c appr},   \\
  \bar{N} &= \frac{\rho t v_{\rm{enc}}}{\pi}, \label{eq-N glo appr} \\
  \bar{V} &= v_{\rm{enc}}, \label{eq-V glo appr} \\
  \bar{V}_\bot &= \frac{\pi v_{\rm{enc}}}{4}, \label{eq-VV glo appr} \\
  \bar{D} &= c_{\rm{c}} \bar{d}_{\rm{p}}^{\gamma'_{\rm{c}}} (\frac{\pi v_{\rm{enc}}}{4})^{\gamma_{\rm{c}}}, \label{eq-D glo appr} \\
  \bar{N}_{\rm{c}} &= \frac{\rho t v_{\rm{enc}}}{\pi} [\frac{d_{\rm{min}}}{c_{\rm{p}} d_{\rm{c}}^{\gamma'_{\rm{p}}}} (\frac{\pi v_{\rm{enc}}}{4})^{\gamma_{\rm{p}}}]^{\alpha_{\rm{p}}} \label{eq-N_c glo appr}.
\end{align}
All of the approximate distributions are shown in Fig. \ref{fig-beta} with dashed curves for comparison with the exact ones. Smaller $\eta$ leads to better approximation. Except for $N_{\rm{c}}(\beta)$ and $\bar{N}_{\rm{c}}$, the errors of the approximate forms are $\lesssim 1\%$ for $\eta = 0.1$ and $\lesssim 5\%$ even for $\eta =0.25$. In particular, the errors of $V_\bot(\beta)$ are merely 0.04\% and 0.25\%, respectively. The approximations of $N_{\rm{c}}(\beta)$ and $\bar{N}_{\rm{c}}$ also depend on $\alpha_{\rm{p}}$. Given $\eta = 0.1$, their errors are 1\%--5\% and 0.7\%--1.6\% with $\alpha_{\rm{p}}$ increasing from 1 to 3, which is acceptable. However, for extreme conditions when $\eta$ and $\alpha_{\rm{p}}$ are both large, approximate forms of $N_{\rm{c}}(\beta)$ and $\bar{N}_{\rm{c}}$ should be used with caution. For reference, the current Earth-Moon distance $a_{\rm{M}} = 60 R_{\oplus}$ ($v_{\rm{M}} = 1.0$ km~s$^{-1}$) and the current impactor population of NEOs ($v_{\rm{enc}} \simeq 20$ km~s$^{-1}$ according to \citet{Gallant2009}) result in an $\eta$ of only about 0.05.

There are a few points about the approximate distributions to note. First, all the spatial distributions can be described by a simple function of $\beta$,
\begin{equation}
    \Gamma(\beta) = A_0 (1 + A_1 \cos\beta),
\end{equation}
where $A_0 > 0$ and $0 < A_1 < 1$. This function is symmetric about the point $(90\degr, A_0)$ and has one maximum $A_0 (1 + A_1)$ at the apex ($\beta = 0\degr$) and one minimum $A_0 (1 - A_1)$ at the antapex ($\beta = 180\degr$), with a monotonic decrease from the former to the latter. Thus, the leading/trailing asymmetry is again proved here through analysis.

Second, the two parameters $A_{0}$ and $A_1$ entirely determine the variation of $\Gamma$. It is true that for every $\Gamma$, $A_{0}$ is exactly the approximate $\bar{\Gamma}$ and also the value of $\Gamma$ on the prime meridian ($\beta = 90\degr$). Additionally, $A_0$ has a positive relation with $v_{\rm{enc}}$ (except for $\Theta$), explaining the dependence of $\bar{\Gamma}$ on $v_{\rm{enc}}$ shown in Fig. \ref{fig-lam}. (The effect of $\eta$ on the exact $\bar{\Gamma}$ is negligible.) The factor $(1+A_1\cos\beta)$ is then equal to the normalized distribution $\Delta\Gamma$. The second parameter $A_1$ just determines the amplitude of the relative variation shown in Fig. \ref{fig-beta} and therefore is called "asymmetry amplitude". For all $\Gamma$, it always holds that $A_1 \propto \eta$, that is, the higher the $v_{\rm{enc}}$ or the lower the $v_{\rm{M}}$ (the larger $a_{\rm{M}}$), the fainter the leading/trailing asymmetry. This relation can be understood by imaging the movement of the normal impact point along the lunar equator, whose rate is ${\rm{d}}\lambda_\bot/{\rm{d}}t$. If it were moving uniformly, that is, if $\eta \rightarrow 0$ and thus ${\rm{d}}\lambda_\bot/{\rm{d}}t \rightarrow -{\rm{d}}f_{\rm{M}}/{\rm{d}}t = -n_{\rm{M}}$, then the conditions (impact number, size distribution of craters, etc.) in every unit area would be the same, leading to a uniform spatial distribution. The varying moving rate adds to the number of normal impacts near the apex, and thus results in the biased cratering in all aspects. In this way, $\eta$ determines $A_1$.

Third, some simple relations between $A_{0,1}$ of different variables are found. Parameters $A_{0,1}$ of $N$, $V$, $\Theta$, $V_\bot$, $D$, and $N_{\rm{c}}$ are denoted with superscripts $N$, $V$, $\Theta$, $\bot$, $D$, and c, respectively. It is seen that
\begin{align}
  A_0^\bot &= A_0^V A_0^\Theta,    \label{eq-A0VV rel}  \\
  A_0^D &= c_{\rm{c}} \bar{d}_{\rm{p}}^{\gamma'_{\rm{c}}} (A_0^\bot)^{\gamma_{\rm{c}}},    \label{eq-A0D rel} \\
  A_0^{\rm{c}} &= (\frac{d_{\rm{min}}}{c_{\rm{p}} d_{\rm{c}}^{\gamma'_{\rm{p}}}})^{\alpha_{\rm{p}}} (A_0^\bot)^{\gamma_{\rm{p}} \alpha_{\rm{p}}} A_0^N ,   \label{eq-A0c rel}\\
  A_1^D &= \gamma_{\rm{c}} A_1^\bot,   \label{eq-A1D rel}  \\
  A_1^{\rm{c}} &= \gamma_{\rm{p}} \alpha_{\rm{p}} A_1^\bot + A_1^N. \label{eq-A1c rel}
\end{align}
These are the results of the relations between the variables, and the first three equations involving $A_0$, the approximate $\bar{\Gamma}$, are identical to the connections between $V$, $\Theta$, $V_\bot$, $D$, and $N_{\rm{c}}$ themselves (Eqs. \ref{eq-D lambda} and \ref{eq-N_c lambda}). Furthermore, the last two relations between $A_1$, the asymmetry amplitudes, are degenerate relations between $A_0$: the product of $(A^\bot)^{\gamma_{\rm{p}}\alpha_{\rm{p}}}$ and $A^N$ becomes their sum, and then the exponents of $A^\bot$, both $\gamma_{\rm{c}}$ and $\gamma_{\rm{p}}\alpha_{\rm{p}}$, become its coefficients. The absolute distributions of $D$ and $N_{\rm{c}}$ both depend on the impactors' size distribution, which is characterized by $\alpha_{\rm{p}}$ and $d_{\rm{min}}$, but $A_1^D$ is independent of this, and $A_1^{\rm{c}}$ only involves $\alpha_{\rm{p}}$. That $A_1^{\rm{c}}$ has no dependence on $d_{\rm{c}}$ means that the normalized crater density distribution is not a function of $d_{\rm{c}}$: $N_{\rm{c}}(>d_{\rm{c}},\beta) / \bar{N}_{\rm{c}}(>d_{\rm{c}}) = \Delta N_{\rm{c}}(\beta)$.

Last but not the least, the above facts inspire the observation. Except for $D$ and $N_{\rm{c}}$, the other four, $N$, $V$, $\Theta$, and $V_\bot$, are unobservable in surveys of crater record. Now the dependence of $A_{1}^{D,\rm{c}}$ on $A_{1}^{N,\bot}$ provides a way to directly obtain the relative variations of $N$ and $V_\bot$. Furthermore, when $\eta$ is derived from the normalized distributions of $D$ and $N_{\rm{c}}$, those of all the unobservable variables can be obtained. Since $A_1^{\rm{c}}$ has no dependence on $d_{\rm{c}}$, when observed craters are counted on the Moon, it is better to choose a small $d_{\rm{c}}$ for accuracy based on more data, as long as the exclusion of saturation and secondaries is ensured and $d_{\rm{p}}(d_{\rm{c}},V_\bot(\beta)) > d_{\rm{min}}$ holds (Sect. \ref{subsec-exa}). Conversely, we can also determine whether secondary craters are included in a sample or whether a sample is complete by verifying if $N_{\rm{c}}(>d_{\rm{c}},\beta)/\bar{N}_{\rm{c}}(>d_{\rm{c}})$ is different from when $d_{\rm{c}}$ is greater. It is even more meaningful that formulations $D(\beta)$ and $N_{\rm{c}}(>d_{\rm{c}},\beta)$ enable us to derive from the crater record the bombardment conditions, which includes not only the impactor properties and bombardment duration, but also the lunar orbit in the dominant epoch of the impactors (Sect. \ref{subsec-rep}).

\subsection{Reproducing bombardment conditions}   \label{subsec-rep}

\begin{figure}
    \resizebox{\hsize}{!}{\includegraphics{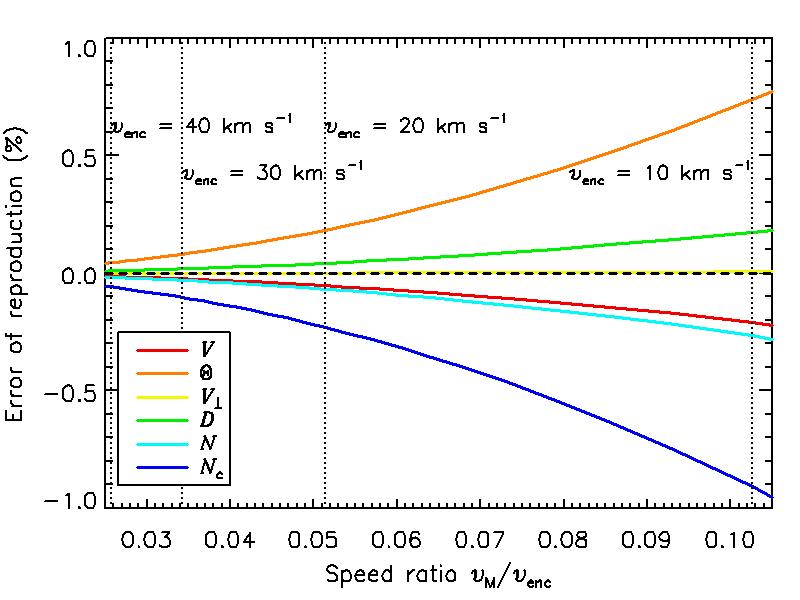}}
    \caption{Reproduction error of the speed ratio $\eta$. For a given exact $\eta$, after fitting the approximate formulations to the exact cratering distribution (assuming $\alpha_{\rm{p}}$ = 1.75) and using the theoretical relations between asymmetry amplitude and speed ratio to reproduce the latter with the fits of the former, the reproduction error is calculated as the relative difference between the reproduced and the exact $\eta$. The reproduction errors that the spatial distributions of $V$, $\Theta$, $V_\bot$, $D$, $N$, and $N_{\rm{c}}$ lead to (red, orange, yellow, green, cyan, and blue curves) are compared. Where $v_{\rm{enc}}$ = 10, 20, 30, and 40 km~s$^{-1}$ on condition $v_{\rm{M}} = 1.0$ km~s$^{-1}$ are indicated (vertical dot lines).}
    \label{fig-rep}
\end{figure}

To verify the validity of the approximate formulations (Eqs. \ref{eq-N appr}--\ref{eq-N_c appr}), we fit them using the least-squares method to the exact distributions (Eqs. \ref{eq-N beta}--\ref{eq-N_c beta}) that we considered as observed data. The fit cratering distributions are shown in Fig. \ref{fig-beta} with dotted curves. Given $\alpha_{\rm{p}}$, which only influences the relative variation of $N_{\rm{c}}$, it is seen that the smaller the $\eta$, the smaller the difference between the fit and exact distributions, because smaller $\eta$ leads to a better approximation at first. The errors of fit distributions are $<0.3\%$ when $\eta = 0.1$ and $\alpha_{\rm{p}} = 1.75$, almost half the errors of the approximate ones, except that for $\Theta$ they are nearly equal. The best fits are for the $V_\bot$ distribution, with errors $<0.02\%$.
Furthermore, given the fits of $A_{1}$, $\eta$ can be easily reproduced based on the relations between $A_1$ and $\eta$. Figure \ref{fig-rep} shows how the reproduced $\eta$ varies with the exact $\eta$. When $\eta = 0.1$ and $\alpha_{\rm{p}} = 1.75$, the errors of the reproduced $\eta$ are lower than 1\% for all the variables. As expected, smaller $\eta$ leads to a better reproduction of itself. Moreover, fitting distributions of different variables results in different reproductions: $\Theta$ and $N_{\rm{c}}$ (given $\alpha_{\rm{p}} = 1.75$) give relatively large errors, while $V_\bot$ gives the smallest; $\Theta$ and $D$ tend to overestimate, while $N$, $V$, and $N_{\rm{c}}$ tend to underestimate.
We caution that the observational error and limited data will prevent reaching such good fits and reproductions in reality, but our estimation shows what the series of approximate formulations is capable of.

The problem is that of $N$, $V$, $\Theta$, $V_\bot$, $D$, and $N_{\rm{c}}$, the first four are not observables of formed craters (except elliptic rim of a crater may imply extreme oblique impact). We therefore propose a method for reproducing the bombardment conditions (the lunar orbit and its impactor population) when a given crater record was formed through combined observations of $D$ and $N_{\rm{c}}$. By fitting formulations $D(\beta)$ and $N_{\rm{c}}(\beta)$ to the observed spacial distributions of $D$ and $N_{\rm{c}}$ and then solving the system of equations,
\begin{align}
  A_0^D &= c_{\rm{c}} \bar{d}_{\rm{p}}^{\gamma'_{\rm{c}}} (\frac{\pi v_{\rm{enc}}}{4})^{\gamma_{\rm{c}}}, \label{eq-A0D}  \\
  A_0^{\rm{c}} &= \frac{F t}{l_{\rm{gl}}} [\frac{d_{\rm{min}}}{c_{\rm{p}} d_{\rm{c}}^{\gamma'_{\rm{p}}}} (\frac{\pi v_{\rm{enc}}}{4})^{\gamma_{\rm{p}}}]^{\alpha_{\rm{p}}},  \label{eq-A0c} \\
  A_1^D &= \frac{(16-\pi^2)\gamma_{\rm{c}}}{2\pi} \frac{v_{\rm{M}}}{v_{\rm{enc}}}, \label{eq-A1D} \\
  A_1^{\rm{c}} &= \frac{\pi^2 + (16-\pi^2)\gamma_{\rm{p}} \alpha_{\rm{p}}}{2\pi} \frac{v_{\rm{M}}}{v_{\rm{enc}}} \label{eq-A1c},
\end{align}
where the left-hand sides are fits of $A_{0,1}^{D}$ and $A_{0,1}^{\rm{c}}$, we may reproduce the complete set of bombardment conditions: the lunar orbital speed $v_{\rm{M}}$ and the impactors' encounter speed $v_{\rm{enc}}$, slope of the size distribution $\alpha_{\rm{p}}$, minimum diameter $d_{\rm{min}}$, impact flux $F$, and dominant duration $t$. We note that $F = 2 R_{\rm{M}} \rho v_{\rm{enc}}$ is used instead of $\rho$ only for convenience in observation, that $\bar{d}_{\rm{p}} = \bar{d}_{\rm{p}}(\alpha_{\rm{p}},d_{\rm{min}})$ (Eq. \ref{eq-d_p mean}), and that $c_{\rm{c,p}}$, $\gamma_{\rm{c,p}}$, $\gamma'_{\rm{c,p}}$, (Eqs. \ref{eq-c_p}--\ref{eq-gamma'_c}) $l_{\rm{gl}}$, and $d_{\rm{c}}$ are all constants. First, even without any knowledge of the above conditions, $\alpha_{\rm{p}}$ and $\eta = v_{\rm{M}}/v_{\rm{enc}}$ can be directly derived from Eqs. \ref{eq-A1D} and \ref{eq-A1c}. Second, with $\alpha_{\rm{p}}$ and $\eta$ obtained, knowing any of $v_{\rm{M}}$, $v_{\rm{enc}}$, $d_{\rm{min}}$, and $F t$ results in acquirement of all of them, because $v_{\rm{M}}$ and $v_{\rm{enc}}$ are related by $\eta$, while $v_{\rm{enc}}$, $d_{\rm{min}}$, and $F t$ are related through Eqs. \ref{eq-A0D} and \ref{eq-A0c}. Third, with $F t$ obtained, knowing either $t$ or $F$ leads to the reproduction of the other.

We now show a sample reproduction following the above procedure. The exact distributions of $D$ and $N_{\rm{c}}$ are taken again as the observed ones under current conditions where the lunar impactor population are the NEOs \citep{Strom2005}: $v_{\rm{M}} = 1.022$ km~s$^{-1}$, $v_{\rm{enc}} = 20$ km~s$^{-1}$ \citep{Gallant2009}, $\alpha_{\rm{p}} = 1.75$ \citep{Bottke2002}, $d_{\rm{min}} = 0.5$ km, $F = 7.5 \times 10^{-13}$ yr$^{-1}$ (for craters larger than 1 km; \citealp{LeFeuvre2011}), and $t = 3.7$ Gyr (age of the Orientale basin; \citealp{LeFeuvre2011}). The value of $d_{\rm{min}}$ and $d_{\rm{c}} = 25$ km are chosen for consistency with the working range of Eqs. \ref{eq-d_c} and \ref{eq-d_p}. We note that our synthetic distributions do not represent the real observation because the real minimum size of NEOs is smaller than what we assume here, and the impact flux is defined to take all the craters into account and not only craters larger than 1 km. However, the goodness of reproduction are not affected.
Fitting the synthetic distributions that each involve 181 data points (uniform $\beta$ interval of $1\degr$) results in $A_0^D = 33.0$ km, $A_1^D = 0.0238$, $A_0^{\rm{c}} = 1.63 \times 10^{-7} \rm{km}^{-1}$, and $A_1^{\rm{c}} = 0.129$ (with uncertainties $\lesssim 0.1\%$), among which $A_1^{D}$ and $A_1^{\rm{c}}$ lead to $\eta = (5.112 \pm 0.003) \times 10^{-2}$ and $\alpha_{\rm{p}} = 1.738 \pm 0.003$. If $v_{\rm{enc}}$ is known, then $v_{\rm{M}} = 1.0224 \pm 0.0006$ km~s$^{-1}$, $d_{\rm{min}} = 0.495 \pm 0.004$ km, and $Ft = (2.82 \pm 0.04) \times 10^{-3}$. Furthermore, if $t$ is given, then $F = (7.6 \pm 0.1) \times 10^{-13}$ yr$^{-1}$. The reproduction errors, differences between the best-fit and the exact values, are 0.05\%, 0.16\%, 0.82\%, and 1.44\% for $v_{\rm{M}}$, $\alpha_{\rm{p}}$, $d_{\rm{min}}$, and $F$, respectively. The goodness of reproduction sufficiently exhibits the viability of the reproducing method.

\section{Numerical simulations} \label{sec-simul}

As indicated by \citet{Strom2015}, the contribution of MBAs to the crater density exceeds that of NEOs by more than an order of magnitude for craters larger than 10 km in diameter, therefore it is quite important to determine the consequence of the bombardment of MBAs. We numerically simulated this process in the background of the LHB, dominance of the MBAs, with the Earth-Moon distance $a_{\rm{M}}$ varying in different cases. We found both the leading/trailing and the pole/equator cratering asymmetries. We established the formulation of the coupled-asymmetric distribution. In addition, the analytical model and numerical simulations confirm each other in various aspects.

\subsection{Numerical model} \label{subsec-Initial}

Our numerical model includes the Sun, Earth, the Moon, and $3\times10^5$ small particles in each case, which represent the primitive MBAs. All the large bodies were treated as rigid spheres. Only gravitation was involved and the effects between particles were ignored. Efforts were made to produce the circumstances of the LHB and a relatively realistic orbit of the Moon whose eccentricity and inclination are considered, so that the simplification of our analytical model can be assessed. The initialization was set at the moment after the MBAs had been disturbed, so that the giants can be excluded. Still, we caution that the orbits of the MBAs are artificially biased and are not modified by the giant planets.

The initial orbit of Earth was assumed to be the same as today. In the ecliptic coordinate system, its semi-major axis was $a_{\oplus}$ = 1 AU, the eccentricity was $e_{\oplus}$ = 0.0167, the inclination was $i_{\oplus} = 0\degr$, the longitude of the ascending node was $\Omega_{\oplus} = 348.7\degr$, the argument of perihelion was $\omega_{\oplus} = 114.2\degr$, and the mean anomaly was $M_{\oplus} = 0\degr$. Because the lunar early history involves a great uncertainty as mentioned in Sect. \ref{sec-intro}, five cases were set with the lunar semi-major axis $a_{\rm{M}}$ = 20, 30, 40, 50, and 60 $R_{\oplus}$ in turn, probably covering most durations of the lunar history. This also helps to examine the influence of $a_{\rm{M}}$ on the cratering. The lunar eccentricity $e_{\rm{M}}$ was initially set to 0.04 \citep{Ross1989}, while the current value of the lunar inclination to the ecliptic $i_{\rm{M}} = 5\degr$ was adopted. The Moon's longitude of the ascending node $\Omega_{\rm{M}}$, the argument of perihelion $\omega_{\rm{M}}$, and the mean anomaly $M_{\rm{M}}$ were all 0$\degr$, because the lunar orbital period and the period of its precession along the ecliptic are extremely short relative to the timescale of the tidal evolution.

Primitive MBAs mean the asteroids that occupied the region where the contemporary main belt is during the LHB, that is, 2.0 AU $\le a_{\rm{p}} \le$ 3.5 AU. Their orbits were disturbed by the migration of giants \citep{Gomes2005}, while their size distribution can be taken as invariant \citep{Bottke2005}. We only needed those MBAs that have the potential for encountering the Earth-Moon system, namely, those with perihelion distances $q_{\rm{p}}$ no larger than 1 AU. To maximize the impact probability \citep{Morbidelli1998} and avoid the interference of the effect of $q_{\rm{p}}$, the initial $q_{\rm{p}}$ was constrained to be 1 AU, resulting in $0.50 \le e_{\rm{p}} \le 0.71$. The inclination was set to $0\degr \le i_{\rm{p}} \le 0.5\degr$ to examine the mechanism of the latitudinal cratering asymmetry. The orbital elements $\Omega_{\rm{p}}$, $\omega_{\rm{p}}$, and $M_{\rm{p}}$ all ranged from 0 to 360$\degr$. Each particle's initial orbital elements were randomly generated in corresponding ranges following uniform distributions, except that its eccentricity was derived from $e_{\rm{p}} = 1 - q_{\rm{p}}/a_{\rm{p}}$.
The mass of every particle was $4 \times 10^{12}$ kg, equivalent to a diameter $d_{\rm{p}} \sim 1$ km given a density of 3 g~cm$^{-3}$. This means that the size distribution is not considered at the stage of $N$-body simulations. The total mass of the particles of $1 \times 10^{18}$ kg is negligible to that of the Moon, therefore generating a size distribution in the simulations will not lead to statistically different results, but will take many more simulations to derive meaningful statistics. Therefore, we only considered this in the post-processing when the crater diameter and crater density are calculated. Although the size distribution of current MBAs has been studied by several surveys (Sect. \ref{sec-intro}), it cannot be easily modeled because of its power-law breaks. In this work, we assumed the normalized size distribution (Eq. \ref{eq-N_p}) to be a single power-law characterized by $\alpha_{\rm{p}} = 3$ and $d_{\rm{min}} = 5$ km referring to \citet{Ivezic2001}, who determined $\alpha_{\rm{p}}$ = 3 for 5 km $\lesssim d_{\rm{p}} \lesssim$ 40 km.

Three-dimensional $N$-body simulations were run to integrate the orbits of the Sun, Earth, the Moon, and $3\times10^5$ particles in each case. The integration time was 10 Myr as a compromise between computing efficiency and the fact that the LHB lasted for $\sim 0.1$ Gyr \citep{Gomes2005}. In addition, this is long enough to allow nearly all the potential impacts to occur, and it is short enough to omit the tidal evolution.
The integration consists of two steps. We first integrated Earth, the Moon, and all the particles in the heliocentric ecliptic reference frame using the hybrid symplectic integrator in the Mercury software package \citep{Chambers1999}, which was altered to adapt it to our purpose. When a particle had a close encounter with the Moon, the information of Earth, the Moon, and this particle was recorded. Then in the next step, we used the Runge-Kutta-Fehlberg method to integrate the recorded particles and the Earth-Moon system until all the particles had impacted the Moon. In this step, the perturbation of the Sun was ignored since the integration time it took is only $\sim 10$ min, while the lunar orbital period is five days ($a_{\rm{M}} = 20 R_{\oplus}$) at least.

Given that the Moon rotated synchronously during the LHB (Sect. \ref{subsec-assum}), every impact in the simulations was precisely located on the lunar surface. Because the Moon has a longitudinal geometrical libration due to its eccentricity (the lunar spin axis was assumed to be perpendicular to the lunar orbital plane so that there is no libration in latitude), the near side is defined as the hemisphere facing Earth only when the Moon is at its perigee and the far side is its opposite. The longitude line that the center of the near side lies on is the prime meridian, with the east longitude values being positive. The apex point, the center of the leading side, is at $(-90\degr, 0\degr)$, while the antapex point, the center of the trailing side, is at $(90\degr, 0\degr)$. The low- and high-latitude regions are also defined here as the area with a latitude between $\pm45\degr$ and the remaining area on the lunar surface, respectively.

Based on this numerical model, the bombardment conditions are $v_{\rm{enc}}$ = 8.32 km~s$^{-1}$ (estimated with Eq. \ref{eq-v_enc} given $q_{\rm{p}}$ = 1 AU and $a_{\rm{p}}$ = 2.75 AU, the mean of the initial $a_{\rm{p}}$), $\alpha_{\rm{p}} = 3$, $d_{\rm{min}} = 5$ km ($\bar{d}_{\rm{p}} = 7.5$ km assuming that the size distribution of the impactors is globally invariant), $t$ = 10$^7$ yr, and $v_{\rm{M}}$ = 1.78, 1.45, 1.26, 1.12, and 1.03 km~s$^{-1}$ (with $e_{\rm{M}}$ ignored) for cases 1--5. Hereafter the above modeled bombardment conditions and the results we derived from them using our analytical formulations are called "predicted", while those directly derived form simulations are called "simulated".

\subsection{Preliminary statistics} \label{subsec-stat}

\begin{figure*}
    \centering
    \includegraphics[width=17cm]{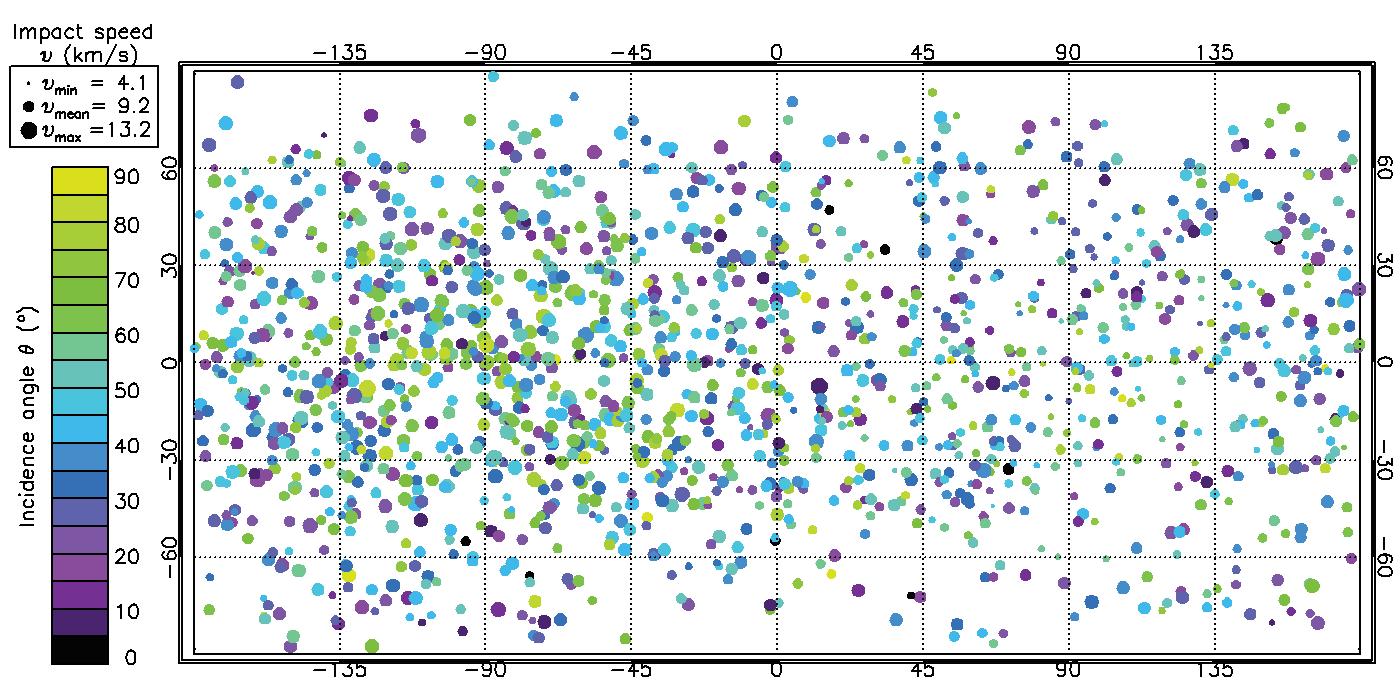}
    \caption{Impact distribution on the lunar surface for the case $a_{\rm{M}}$ = 20 $R_{\oplus}$ with 1544 impacts in total. Every impact in the simulations is shown at the location it occurs with a point, whose size and color represent the magnitudes of impact speed and incidence angle for this impact.}
    \label{fig-imp dis}
\end{figure*}

\begin{figure*}
    \centering
    \includegraphics[width=17cm]{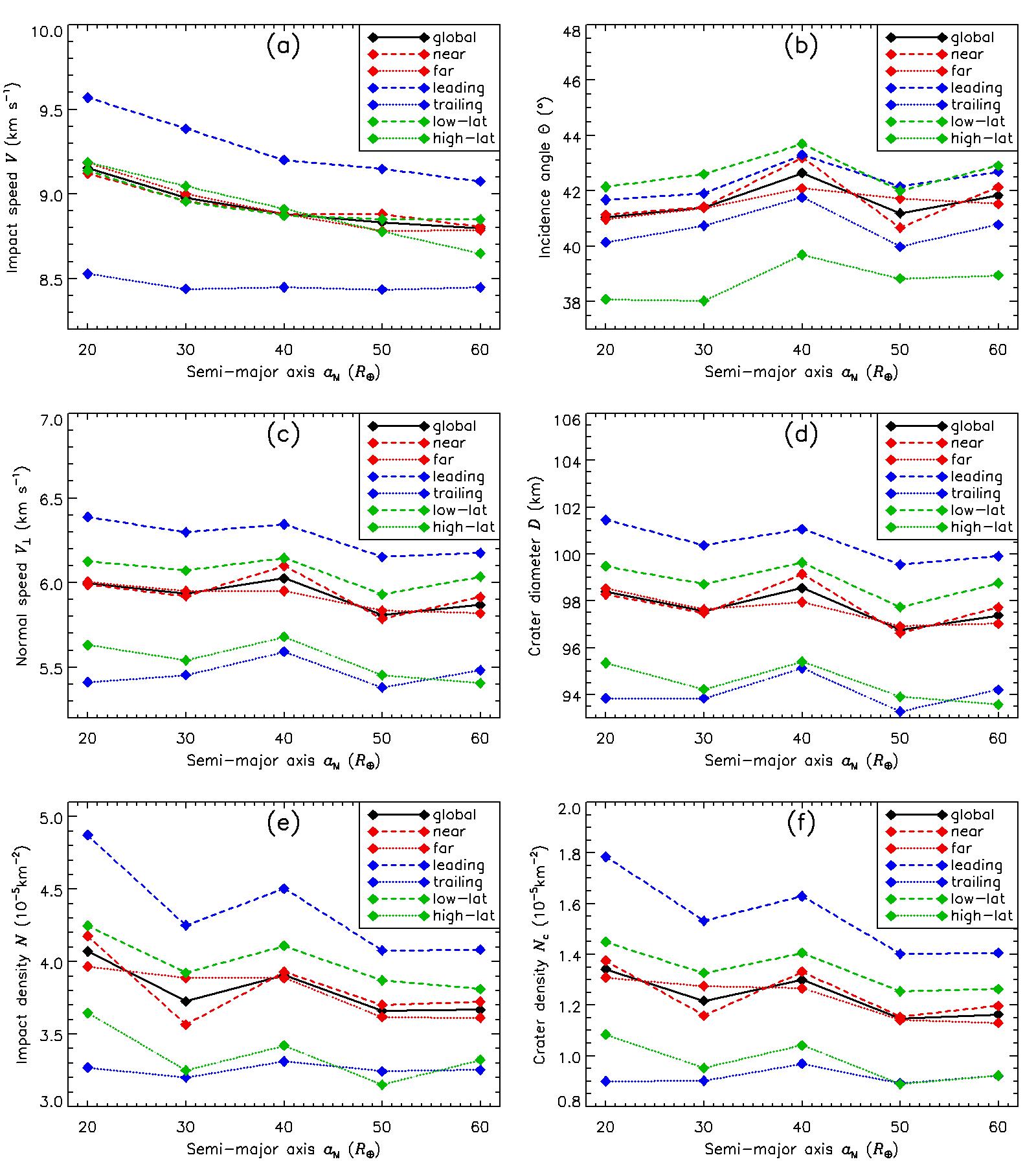}
    \caption{Regional impact speed (a), incidence angle (b), normal speed (c), crater diameter (d), impact density (e), and crater density (f) as functions of the lunar semi-major axis. The black squares connected by black solid lines are the global averages. The red squares connected by the red dashed and dotted lines are the averages on the near and far sides. The blue squares connected by the blue dashed and dotted lines are the averages on the leading and trailing sides. The green squares connected by the green dashed and dotted lines are the averages on the low- and high-latitude regions.}
    \label{fig-Gamma vs EMD}
\end{figure*}

Cases 1--5 with $a_{\rm{M}}$ = 20, 30, 40, 50, and 60 $R_{\oplus}$ in turn result in the global impact number $C_{\rm{gl}}$ = 1544, 1413, 1482, 1388, and 1391, respectively. Figure \ref{fig-imp dis} shows the impact distribution on the lunar surface for case 1, and those of the other cases are similar. Every impact location was recorded along with the impact condition, namely, the impact speed $v$ and the incidence angle $\theta$. After this, the normal speed $v_\bot = v \sin\theta$ and the size of the formed crater centered on this impact location $d_{\rm{c}} = d_{\rm{c}}(\bar{d}_{\rm{p}}, v_\bot)$ were calculated. In a specific area, the averages of $v$, $\theta$, $v_\bot$, and $d_{\rm{c}}$ are denoted by $V$, $\Theta$, $V_\bot$, and $D$, following Sect. \ref{sec-analy}; the impact density $N$ is the number of impacts $C$ divided by the area, while the crater density $N_{\rm{c}}$ is the number of craters larger than a given size $d_{\rm{c}}$ divided by the area. We point out that this crater number was obtained by totalling the probability of generating a crater larger than $d_{\rm{c}}$ for every impact in this area, that is, $\sum_{j=1}^{C} \bar{N}_{\rm{p}}(> d_{\rm{p}}(d_{\rm{c}}, v_{\bot,j}))$. This is how the size distribution of impactors is involved in post-processing. The given size $d_{\rm{c}}$ was set to 100 km to ensure $d_{\rm{p}}(d_{\rm{c}}, v_\bot) \ge d_{\rm{min}}$, otherwise the descriptions of $N_{\rm{c}}$ distribution does not hold (Sect. \ref{subsec-exa}). We note that the assumed values of $\bar{d}_{\rm{p}}$ and $d_{\rm{c}}$ do not influence the relative spatial variations of $D$ and $N_{\rm{c}}$ at all.

We summarize the regional $\Gamma$ (denoting every/any one of $V$, $\Theta$, $V_\bot$, $D$, $N$, and $N_{\rm{c}}$ following Sect. \ref{sec-analy}) for cases 1--5 in Fig. \ref{fig-Gamma vs EMD}. The impact density of the leading side $N_{\rm{ldg}}$ is much higher than that of the trailing side $N_{\rm{trg}}$ for all the cases, with excesses of 25\%--49\% (equivalent to the excesses of $C_{\rm{ldg}}$ to $C_{\rm{trg}}$). Other $\Gamma_{\rm{ldg}}$ also remain larger than $\Gamma_{\rm{trg}}$ with notable excesses. These are clear signs of the leading/trailing asymmetry, indicating its existence in all the investigated aspects of cratering at any $a_{\rm{M}}$. It also shows unambiguously that except for $V$, the averages of the low-latitude region $\Gamma_{\rm{low}}$ are always higher than those of high-latitude region $\Gamma_{\rm{high}}$, which confirms the pole/equator asymmetry regardless of $a_{\rm{M}}$. We note that of the investigated variables, $V$ is the most sensitive to the leading/trailing asymmetry, but does not involve a pole/equator asymmetry at all, while $\Theta$ has the greatest dominance in the pole/equtor asymmetry over the leading/trailing asymmetry. No one pair of $\Gamma_{\rm{near}}$ and $\Gamma_{\rm{far}}$ has a distinct and lasting discrepancy. For example, $N_{\rm{near}}$ is 8\% smaller than $N_{\rm{far}}$ for case 2, but 1\%--5\% greater for other cases (the same for $C_{\rm{near}}$ and $C_{\rm{far}}$). Therefore, the near/far asymmetry is clearly negligible for the whole range 20--60 $R_{\oplus}$ even if it is present, in accordance with \citet{Gallant2009} and \citet{LeFeuvre2011}, who adopted the current impactors.

It is apparent that $\bar{V}$ monotonically decreases from 9.15 to 8.80 km~s$^{-1}$ with $a_{\rm{M}}$ increasing. The predicted $\bar{V}$ (Eq. \ref{eq-V glo}) is 8.60--8.41 km~s$^{-1}$ for $a_{\rm{M}}$ = 20--60 $R_{\oplus}$, only 6\%--4\% smaller than simulated values. Although we consider this already an excellent agreement, we continue to take the gravitational acceleration of the Moon into account, which is ignored in our analytical model, by substituting $v_{\rm{enc}}$ = 8.32 km~s$^{-1}$ with $\sqrt{v_{\rm{enc}}^2 + v_{\rm{esc}}^2} = 8.65$ km~s$^{-1}$, where the lunar escape speed is $v_{\rm{esc}} = 2.38$ km~s$^{-1}$. Then the predicted $\bar{V}$ is recalculated to be 8.92--8.74 km~s$^{-1}$, only 2.5\%--0.7\% smaller than the simulated values. The remaining differences may be due to the ignored acceleration from Earth in addition to the statistical errors. It should be pointed out that the unusually low impact speeds result from the fixed initial perihelion distances of projectiles $q_{\rm{p}}$ = 1 AU. For the whole population of the MBAs without this constraint as well as the NEOs, $v_{\rm{esc}}$ is even more negligible, so that the exclusion of the acceleration in the analytical model is quite acceptable.

Unlike the precise prediction of $\bar{V}$, the simulated $\bar{\Theta}$ are all obviously smaller than the analytical prediction of 51.8$\degr$ (Eq. \ref{eq-Theta glo}). The difference between $\sin{\bar{\Theta}}$ and $\pi/4$ is 13.7\%--16.4\% for the five cases. The reason probably is that our analytical model is planar, describing the variation of $\Theta$ on the equator, while in the three-dimensional simulations, the impacts elsewhere on the lunar surface are all statistically more oblique without isotropic impactors, and thus the simulated $\bar{\Theta}$ are diminished. Since theoretically $\bar{\Theta}$ is constant regardless of how severe the leading/trailing asymmetry is (Sect. \ref{subsec-glo}), and the pole/equtor asymmetry resulting from the concentration of low-inclination impactors should have no dependence on $v_{\rm{M}}$, the simulated $\bar{\Theta}$ were expected to be invariant with $a_{\rm{M}}$. In fact, we do find $\bar{\Theta}$ of all the cases lying in a narrow range 41.1--42.6$\degr$ with a fluctuation of only 2\%.

$\bar{V}_{\bot}$ and $\bar{D}$ show a mild dependence on $a_{\rm{M}}$, because their variations with $a_{\rm{M}}$ are combinations of those of $\bar{V}$ and $\bar{\Theta}$ ($\bar{V}_{\bot} = \bar{V} \sin{\bar{\Theta}}$ and $\bar{D} \propto \bar{V}_{\bot}^{\gamma_{\rm{c}}}$). Owing to the fluctuation of the $\bar{\Theta}$ variation, their negative correlations with $a_{\rm{M}}$ are slightly obscured. The dependence on $\bar{\Theta}$ also causes the simulated $\bar{V}_{\bot}$, 5.81--6.03 km~s$^{-1}$, to be lower by 9.3\%--12.2\% than the predicted values of 6.61--6.75 km~s$^{-1}$ (Eq. \ref{eq-VV glo}); this is also true for the simulated $\bar{D}$, 96.8--98.6 km, which are lower by 6.7\%--8.3\% than the predictions of 105.4--106.5 km (Eq. \ref{eq-D glo}). However, the difference between the simulated $\bar{V}_{\bot}$ and the product of the simulated $\bar{V}$ and $\sin{\bar{\Theta}}$ is no more than 0.3\% for all the cases, and the simulated $\bar{D}$ is lower than $c_{\rm{c}} \bar{d}_{\rm{p}}^{\gamma'_{\rm{c}}} \bar{V}_\bot^{\gamma_{\rm{c}}}$ for only 2.6\% at most. This indicates that although the pole/equator asymmetry in the $\Theta$ distribution leads to an overestimation of $\bar{\Theta}$ by the analytical model, the relationships between the global averages derived in Sect. \ref{subsec-glo} are still well applicable.

In general, simulated $\bar{N}$ shows a negative correlation with $a_{\rm{M}}$ as well. (We note that the relative variation of $\bar{N}$ also represents that of $C_{\rm{gl}}$.) The approximate expression of $\bar{N}$ (Eq. \ref{eq-N glo appr}) indicates, however, that it is proportional to the constant $v_{\rm{enc}}$, while the exact expression involving $v_{\rm{M}}$ (Eq. \ref{eq-N glo}) is consistent with this negative correlation. Of the five sets of data points, case 2 ($a_{\rm{M}}$ = 30 $R_{\oplus}$) seems abnormal, where $\bar{N}$ is oddly deficient and $N_{\rm{near}}$ is clearly smaller than $N_{\rm{far}}$. Assuming that the reason is that the Earth's shielding reduces the impacts on the near side and thus the global impact number, which is supported by the fact that $N_{\rm{far}}$ of this case is larger than case 3 ($a_{\rm{M}}$ = 40 $R_{\oplus}$) as expected, case 1 with its smaller $a_{\rm{M}}$ should have suffered the same effect. However, in case 1, $N_{\rm{near}}$ is even larger than $N_{\rm{far}}$. Therefore, we prefer to consider it as a possibility instead of confirming it as evidence of the near/far asymmetry.
Because $\bar{N}_{\rm{c}} \propto \bar{V}_{\bot}^{\gamma_{\rm{p}}\alpha_{\rm{p}}} \bar{N}$, it is natural to see that the variation of $\bar{N}_{\rm{c}}$ with $a_{\rm{M}}$ combines those of $\bar{V}_{\bot}$ and $\bar{N}$. Despite of the wavy fluctuation, $\bar{N}_{\rm{c}}$ generally decreases as $a_{\rm{M}}$ increases. Additionally, $\bar{N}_{\rm{c}}$ is close to $[d_{\rm{min}} / (c_{\rm{p}} d_{\rm{c}}^{\gamma'_{\rm{p}}})]^{\alpha_{\rm{p}}} \bar{V}_{\bot}^{\gamma_{\rm{p}}\alpha_{\rm{p}}} \bar{N}$ with excess of 8.6\%--9.7\% for all the cases, in agreement with Eq. \ref{eq-N_c glo}.

In short, this preliminary statistics confirms the presence of the leading/trailing asymmetry in spatial distributions of all $\Gamma$ and the pole/equator asymmetry in those of all but $V$. All $\bar{\Gamma}$ but $\bar{\Theta}$ show negative correlations with $a_{\rm{M}}$ as indicated by the analytical model, which precisely predicts the $\bar{V}$ value of all cases and mildly overestimates $\bar{\Theta}$ and thus $\bar{V}_\bot$ and $\bar{D}$, while the correlations between global averages apply well.

\subsection{Spatial distributions} \label{subsec-distr}

\begin{figure*}
    \centering
    \includegraphics[width=17cm]{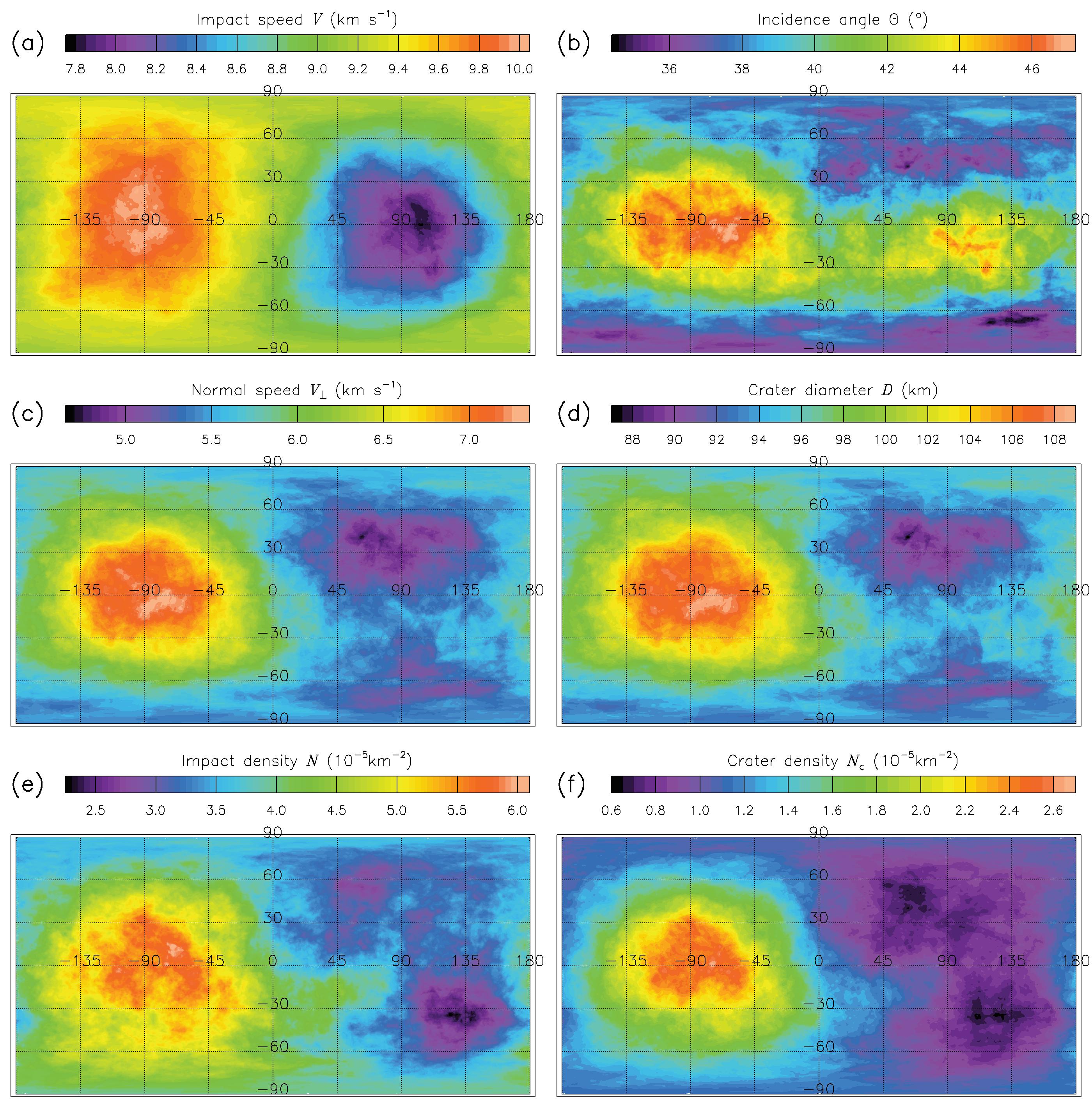}
    \caption{Simulated spatial distribution of impact speed (a), incidence angle (b), normal speed (c), crater diameter (d), impact density (e), and crater density (f) for the case $a_{\rm{M}}$ = 20 $R_{\oplus}$. The value at each location is the average over a circular area centered on this location with a radius of 1061 km, assuming $\bar{d}_{\rm{p}} = 7.5$ km and $d_{\rm{c}} = 100$ km.}
    \label{fig-Gamma sim}
\end{figure*}

\begin{figure*}
    \centering
    \includegraphics[width=17cm]{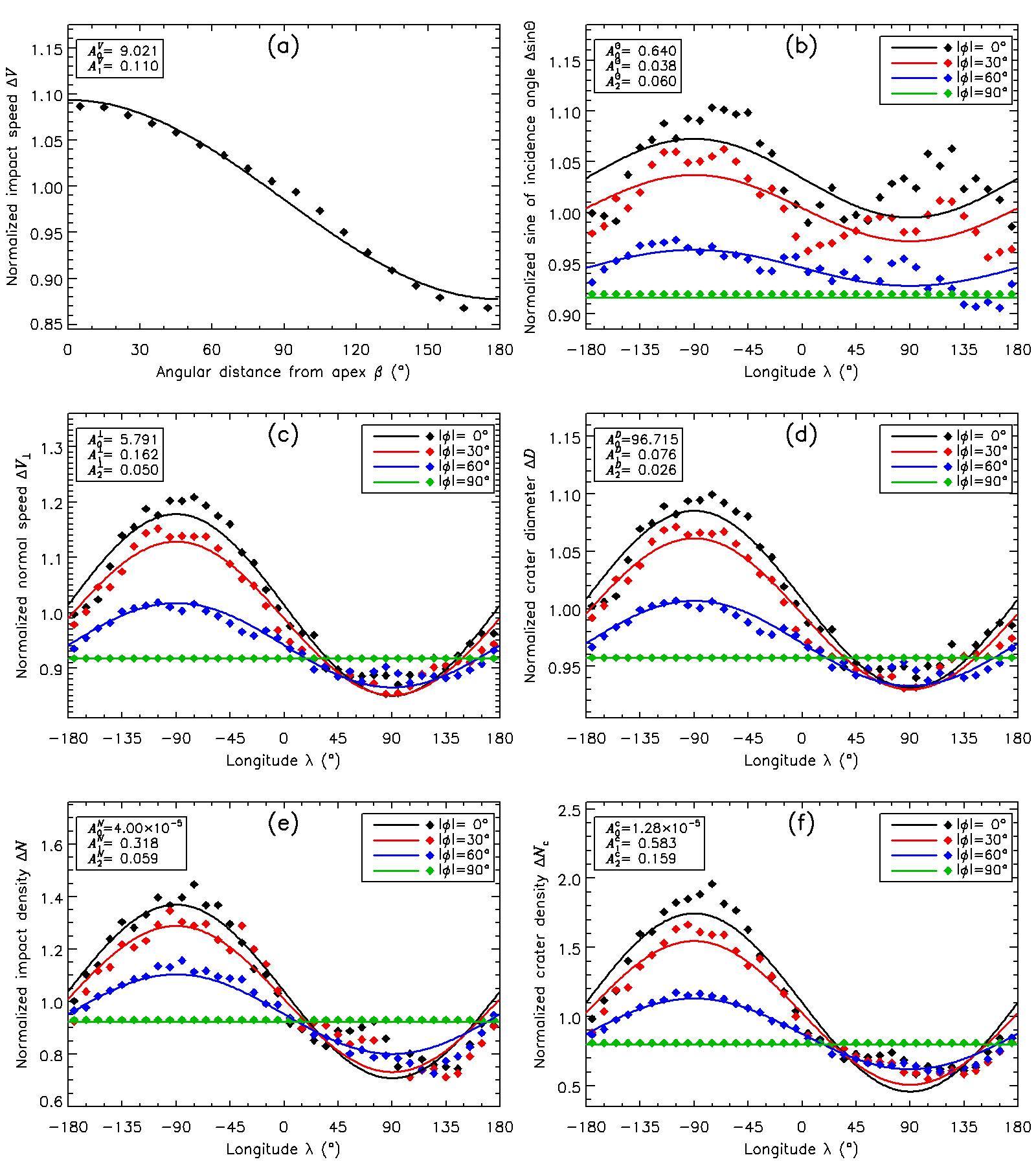}
    \caption{Impact speed as a function of angular distance from the apex (a), and the incidence angle (b), normal speed (c), crater diameter (d), impact density (e), and crater density (f) as functions of longitude, all in terms of simulated global averages. a) Data from the simulated distribution of the impact speed (black squares), each of which is the average over the small circle where $\beta$ is fixed, are fit with Eq. \ref{eq-V fit}. The best fit (black solid curve) is generated with plotted $A_{0,1}^V$. b)--f) Data from simulated distributions, each of which is the mean of the two values at location $(\lambda, \pm\varphi)$ with $|\varphi|$ being 0$\degr$, 30$\degr$, 60$\degr$, and 90$\degr$ (black, red, blue, and green squares), are fit with Eqs. \ref{eq-Theta fit}--\ref{eq-Nc fit}. The best fits (solid curves in the same color as the squares for the same $|\varphi|$) are generated with plotted $A_{0,1,2}$.}
    \label{fig-mullam}
\end{figure*}

The cratering distribution of case $a_{\rm{M}}$ = 20 $R_{\oplus}$ is shown in Fig. \ref{fig-Gamma sim}, and those of other cases have no qualitative differences. Every value of $\Gamma$ on a given location is the average over a circular area centered on this location with a "smooth radius" of $r_{\rm{s}} = 35\degr$ (along a great circle), equivalent to 1061 km on the lunar surface. Smaller $r_{\rm{s}}$ can reveal more details, but because of the limited number of data points, we consider 35$\degr$ as the proper choice for showing important features.

The distribution of $V$ (Fig. \ref{fig-Gamma sim}a) is typical of a pure leading/trailing asymmetry: the maximum 10.0 km~s$^{-1}$ is near the apex, while the minimum 7.8 km~s$^{-1}$ is near the antapex; values on the longitudes of $0\degr$ and $\pm 180\degr$, the dividing line between leading and trailing sides, are approximately equal to the global average 9.2 km~s$^{-1}$; there is an obvious trend for $V$ to be monotonously decreasing from apex to antapex. The distribution of $\Theta$ (Fig. \ref{fig-Gamma sim}b), however, combines properties of both the leading/trailing and pole/equator asymmetries: the maximum 47.2$\degr$ is still near the apex, while the minimum seems to slide from the antapex toward the poles along the longitude $90\degr$ and turns into a pair of valleys on both sides of the equator; the contours around the apex tend to stretch along the latitude lines. Distributions of $V_\bot$, $D$, $N$, and $N_{\rm{c}}$ have similar structures to $\Theta$, but with relatively milder pole/equator asymmetries, which is shown by the closer proximity of their valleys. We describe in Sect. \ref{subsec-couple} how a spatial distribution evolves with increasing leading/trailing and pole/equator asymmetries and show that all the above distributions but that of $V$ are productions of the coupled asymmetries. Again it is not found necessary to involve the near/far asymmetry to explain the variations, as these two hemispheres are roughly identical in all aspects.

As expected, it is found that although $\theta$ of all the impacts lie in an extended range covering 0--90$\degr$ (Fig. \ref{fig-imp dis}), the range of $\Theta$, the average of $\theta$ for every unit area, is quite limited (Fig. \ref{fig-Gamma sim}b). The analysis in Sect. \ref{subsec-imp} indicates that as the normal impact point moves on the lunar surface, on every unit area, impacts from normal to extremely oblique can all occur, and that the varying moving rate adds the normal impacts on the apex, resulting in the asymmetry. In other words, the enlargement near the apex is not caused by the general upward shift of $\theta$, but is due to the increased relative fraction of great $\theta$, which is now verified by simulations.

In addition to the nonuniform moving rate of the normal impact point, the leading/trailing asymmetry can be understood in this way: the lunar velocity adds to the impactors heading toward the Moon and subtracts from the impactors chasing it, so that $N$ is higher near the apex; the subtraction of the lunar velocity from that of the impactors, how the impact velocities are derived, increases the velocity components along the apex-antapex axis of the heading impactors, but decreases those of the chasing impactors, meaning that statistically, not only the magnitudes of the impact velocities are greater near the apex, but their directions are also closer to the apex-antapex axis, so that $V$ and $\Theta$ are also biased. The pole/equator asymmetry is triggered by the low relative inclinations between the lunar equator and projectile orbits in our simulations. This initialization limits the number of impacts arriving at the poles, especially the normal impacts, and thus decreases $N$ and $\Theta$ there. Although near the poles there are fewer impacts and they are statistically more oblique, the magnitudes of the impact velocities are not influenced, so that the pole/equator asymmetry is absent in the $V$ distribution. Since $v_\bot$ and $d_{\rm{c}}$ depend on both $v$ and $\theta$, the distributions of $V_\bot$ and $D$ naturally introduce both asymmetries in themselves. Similarly, because $N_{\rm{c}}$ depends on $V_\bot$ and $N$, whose increases provide both a greater probability of generating large craters and more craters in total, the $N_{\rm{c}}$ distribution also shows the coupled asymmetries.

To describe the pure leading/trailing asymmetry, we use the approximate formulations derived in Sect. \ref{subsec-appr}, whose common form is $\Gamma \propto (1 + A_1 \cos\beta)$, where in a three-dimensional model the angular distance from the apex $\beta$ is related to the longitude $\lambda$ and latitude $\varphi$ by
\begin{equation}    \label{eq-beta}
  \cos\beta = -\sin\lambda \cos\varphi.
\end{equation}
To describe the pure pole/equator asymmetry, we suggest an empirical function
\begin{equation}    \label{eq-pure p/e}
    \Gamma \propto (1 + A_2 \cos2\varphi)
\end{equation}
with a monotonic decrease from the equator to the poles. A spatial distribution of the coupled asymmetries is thus assumed to be
\begin{equation}   \label{eq-couple}
  \Gamma(\lambda, \varphi) = A_0 (1 + A_1 \cos\beta) (1 + A_2 \cos2\varphi).
\end{equation}
It can be easily proved by integrating the right-hand side over the spherical surface that $A_0 = \bar{\Gamma}$ still holds as Sect. \ref{subsec-appr} indicates. At the same time, $A_0$ is the value of $\Gamma$ at location $(0\degr, \pm45\degr)$ or $(180\degr, \pm45\degr)$. The parameters $A_1$ and $A_2$ are leading/trailing and pole/equator asymmetry amplitudes, respectively. They are measurements of the asymmetry degrees, and are assumed to be independent of each other. A detailed discussion of the coupled asymmetries can be found in Sect. \ref{subsec-couple}. We note that because of $r_{\rm{s}}$, the asymmetry amplitudes are reduced. The decrease factor of $A_1$ is estimated to be $\sin r_{\rm{s}}/r_{\rm{s}}$, meaning that the fits of $A_1$ here are the products of this factor and $A_1$ given in Sect. \ref{subsec-appr}. For $r_{\rm{s}} = 35\degr$, this decrease factor is 0.939.

We fit these formulations with the nonlinear least-squares method to the simulated spatial distributions shown in Fig. \ref{fig-Gamma sim}:
\begin{align}
  V(\beta) &= A_0^V (1 + A_1^V \cos\beta),     \label{eq-V fit}    \\
  \sin\Theta(\lambda,\varphi) &= A_0^{\Theta} (1+A_1^{\Theta}\cos\beta) (1+A_2^{\Theta}\cos2\varphi), \label{eq-Theta fit} \\
  V_\bot(\lambda,\varphi) &= A_0^{\bot} (1 + A_1^{\bot} \cos\beta) (1+A_2^{\bot}\cos2\varphi), \label{eq-VV fit} \\
  D(\lambda,\varphi) &= A_0^{D} (1 + A_1^{D} \cos\beta) (1+A_2^{D}\cos2\varphi), \label{eq-D fit}    \\
  N(\lambda, \varphi) &= A^N_0 (1 + A^N_1 \cos\beta) (1 + A^N_2\cos2\varphi),    \label{eq-N fit}    \\
  N_{\rm{c}}(\lambda, \varphi) &= A^{\rm{c}}_0 (1 + A^{\rm{c}}_1 \cos\beta) (1 + A^{\rm{c}}_2\cos2\varphi).  \label{eq-Nc fit}
\end{align}
Figure \ref{fig-mullam} shows the agreement between the simulated and fit variations of the above variables, with the fits of $A_{0,1,2}$ plotted. For a fixed latitude, the variables always decrease with increasing distance from longitude $-90\degr$ toward $+90\degr$, with their variation amplitudes and vertical positions dependent on the latitude. Given fits of $A_{0,1,2}$, the above formulations can lead to fit spatial distributions whose configurations are determined by the relative magnitudes of $A_{1}$ and $A_2$ (Sect. \ref{subsec-couple}).

\subsection{Fit parameters and reproduced conditions}

\begin{figure*}
    \centering
    \includegraphics[width=17cm]{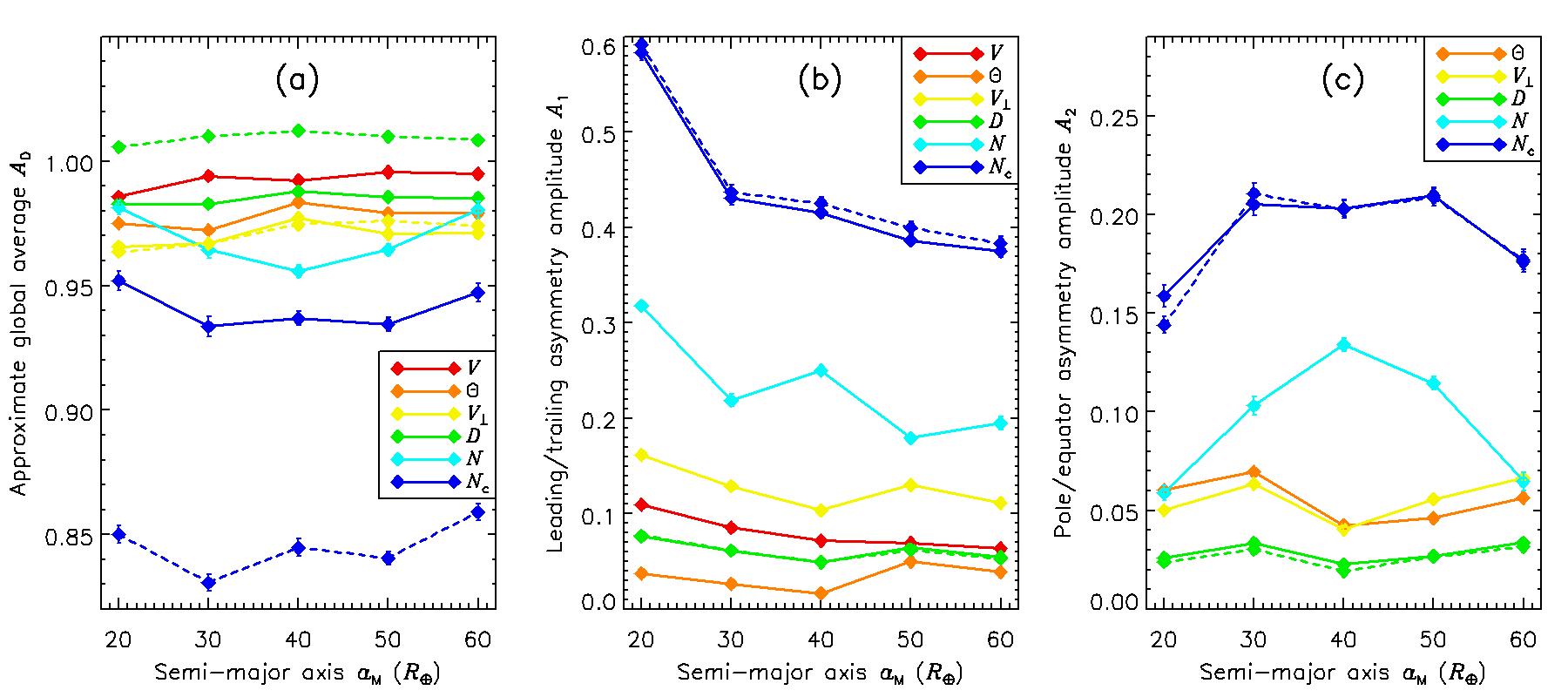}
    \caption{Fit $A_{0}$ (a), $A_1$ (b), and $A_2$ (c) of spatial distributions of impact speed, incidence angle, normal speed, crater diameter, impact density, and crater density (squares connected by solid lines in red, orange, yellow, green, cyan, and blue) as functions of lunar semi-major axis. a) Every approximate global average $A_0$ is in terms of the relevant simulated global average. $A_0^{\bot}$, $A_0^D$, and $A_0^{\rm{c}}$ are compared to $A_0^V A_0^{\Theta}$, $c_{\rm{c}} \bar{d}_{\rm{p}}^{\gamma'_{\rm{c}}} (A_0^\bot)^{\gamma_{\rm{c}}}$, and $[d_{\rm{min}} / (c_{\rm{p}} d_{\rm{c}}^{\gamma'_{\rm{p}}})]^{\alpha_{\rm{p}}} (A_0^\bot)^{\gamma_{\rm{p}} \alpha_{\rm{p}}} A_0^N$ (squares connected by dashed lines in yellow, green, and blue). b) and c) $A_{1,2}^D$ and $A_{1,2}^{\rm{c}}$ are compared to $\gamma_{\rm{c}} A_{1,2}^\bot$ and $\gamma_{\rm{p}} \alpha_{\rm{p}} A_{1,2}^\bot + A_{1,2}^N$ (squares connected by dashed lines in green and blue).}
    \label{fig-A vs EMD}
\end{figure*}

\begin{figure}
    \resizebox{\hsize}{!}{\includegraphics{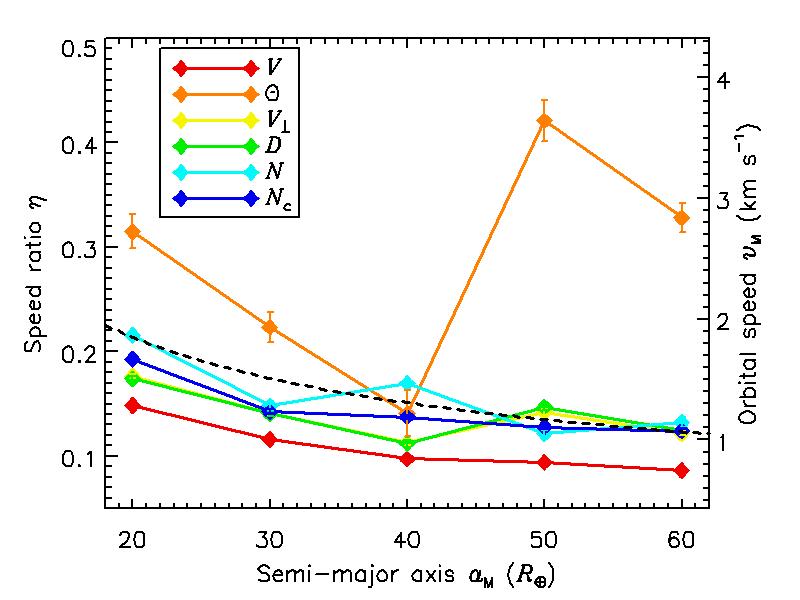}}
    \caption{Speed ratios and lunar orbital speeds reproduced from fit leading/trailing asymmetry amplitudes of impact speed, incidence angle, normal speed, crater diameter, impact density, and crater density (squares connected by solid lines in red, orange, yellow, green, cyan, and blue) in comparison with predictions (black dashed curve).}
    \label{fig-eta vs EMD}
\end{figure}

The qualitative properties of the cratering distribution of case $a_{\rm{M}}$ = 20 $R_{\oplus}$ shown above are common to any other cases. We fit the cratering distribution with Eqs. \ref{eq-V fit}--\ref{eq-Nc fit} for each simulation case. In this subsection, the fit parameters, their quantitative interrelations, and their variations with $a_{\rm{M}}$ are shown and compared with the analytical predictions. Then the bombardment conditions are reproduced.

Fits of $A_0$ of each $\Gamma$ are plotted in terms of simulated $\bar{\Gamma}$ in Fig. \ref{fig-A vs EMD}a. Theoretically, $A_0 \approx \bar{\Gamma}$ regardless of the pole/equator asymmetry (Sect. \ref{subsec-distr}). In Fig. \ref{fig-A vs EMD}a, the fit $A_0$ are all slightly smaller than the simulated $\bar{\Gamma}$, but only by a few percents, which is partly due to the approximation process itself. The largest difference is no more than 7\% between $A^{\rm{c}}_0$ and $\bar{N}_{\rm{c}}$, and the smallest is only 1\% between $A_0^V$ and $\bar{V}$. Therefore, even though we did not rigorously prove that the pole/quator asymmetry can be formulated as Eq. \ref{eq-pure p/e}, Eq. \ref{eq-couple} is clearly a good description of the coupled-asymmetric distribution and the fit of $A_0$ is an excellent reproduction of $\bar{\Gamma}$. The reproduction goodness was expected to be proportional to $\sqrt{a_{\rm{M}}}$ (inversely proportional to $v_{\rm{M}}$) with a given impactor population, but this trend is not observed because it is obscured by the statistical fluctuation. The variation of $\bar{\Gamma}$ with $a_{\rm{M}}$ has been discussed in Sect. \ref{subsec-stat}.
Additionally, the interrelations between $A_0$ (Eqs. \ref{eq-A0VV rel}--\ref{eq-A0c rel}) are verified regardless of the involved pole/equator asymmetry: for any simulation case, $A_0^V A_0^{\Theta}$ approximates to $A_0^{\bot}$ with a difference of $\lesssim0.5\%$; $c_{\rm{c}} \bar{d}_{\rm{p}}^{\gamma'_{\rm{c}}} (A_0^\bot)^{\gamma_{\rm{c}}}$ is almost equivalent to $A_0^D$ with a difference of $<3\%$; and $[d_{\rm{min}} / (c_{\rm{p}} d_{\rm{c}}^{\gamma'_{\rm{p}}})]^{\alpha_{\rm{p}}} (A_0^\bot)^{\gamma_{\rm{p}} \alpha_{\rm{p}}} A_0^N$ is close to $A_0^{\rm{c}}$ with a difference of $\lesssim11\%$.

Our analytical model indicates that $A_1 \propto v_{\rm{M}}/v_{\rm{enc}}$, that is, $A_1 \propto a_{\rm{M}}^{-1/2}$ with a given impactor population. As shown in Fig. \ref{fig-A vs EMD}b, the variation of fit $A_1^V$ with $a_{\rm{M}}$ reflects this relation very well: $A_1^V (a_{\rm{M}}/R_{\oplus})^{1/2}$ = $0.490 \pm 0.010$, $0.469 \pm 0.008$, $0.455 \pm 0.013$, $0.491 \pm 0.004$, and $0.493 \pm 0.006$ for case 1--5 in turn, which is almost a constant. Because the leading/trailing asymmetry in the $\Theta$ distribution is the faintest, the fit $A_1^{\Theta}$ involves relatively great fluctuation, and its relation with $a_{\rm{M}}$ is not clearly seen. The distributions of $V_\bot$ and $D$ depend on those of both $V$ and $\Theta$, so that the fit $A_1^{\bot}$ and $A_1^D$ correlate mildly negatively with $a_{\rm{M}}$. The wavy decrease of fit $A_1^N$ and the monotonic decrease of $A_1^{\rm{c}}$ are also qualitatively consistent with the above theoretic relation.
The interrelations between $A_1$ (Eqs. \ref{eq-A1D rel} and \ref{eq-A1c rel}) also apply quite well. For all cases, the differences between $\gamma_{\rm{c}} A_1^\bot$ and $A_1^D$ and those between $\gamma_{\rm{p}} \alpha_{\rm{p}} A_1^\bot + A_1^N$ and $A_1^{\rm{c}}$ are $\lesssim3\%$.

We have assumed that $A_2$ is invariant with the lunar orbit (Sect. \ref{subsec-distr}). As a result, no common dependence of fit $A_2$ on $a_{\rm{M}}$ is seen in Fig. \ref{fig-A vs EMD}c. To determine whether the assumption is realistic or if the relation is too weak to show itself has to wait for further investigation in the future.
However, referring to the interrelations between the parameters $A_{0,1}$, we find that similar rules are present for $A_2$:
\begin{align}
  A_2^D &= \gamma_{\rm{c}} A_2^\bot,   \label{eq-A2D rel}  \\
  A_2^{\rm{c}} &= \gamma_{\rm{p}} \alpha_{\rm{p}} A_2^\bot + A_2^N. \label{eq-A2c rel}
\end{align}
The comparisons between the left and right sides are also shown in Fig. \ref{fig-A vs EMD}c. We caution that Eqs. \ref{eq-A2D rel} and \ref{eq-A2c rel} are empirically derived and not guaranteed to be valid for other bombardment conditions, although they probably are.

Given the fit $A_1$, the speed ratio $\eta$ and thus the lunar orbital speed $v_{\rm{M}}$ were easily reproduced according to Eqs. \ref{eq-N appr}--\ref{eq-N_c appr}. Figure \ref{fig-eta vs EMD} shows the reproduced $\eta$ and $v_{\rm{M}}$ in comparison with their predictions. The decrease factor was considered in the reproduction. Generally speaking, the reproductions are reasonably consistent with each other. Still, $\eta$ reproduced from $A_1^V$ is obviously smaller than its prediction. Theoretically, $\eta = A_1^V/(\pi/4)$, but the constant quotient of the fit $A_1^V$ and predicted $\eta$ is about 0.53 (with the decrease factor considered), which is smaller than the coefficient $\pi/4$ derived in Sect. \ref{subsec-appr}, so that the reproduced $\eta$ is decreased. Although the reproduction from $A_1^V$ is meant to be lower (Sect. \ref{subsec-rep}), it is not enough to explain the difference of 32\%. The reason why the fit $A_1^V$ precisely follows $A_1^V \propto a_{\rm{M}}^{-1/2}$ but underestimates the predicted $\eta$ and $v_{\rm{M}}$ may be that the leading/trailing asymmetry of the $V$ distribution is decreased by the introduction of the pole/equator asymmetry in the three-dimensional numerical model, so that the coefficient decreases from $\pi/4$ to somewhere lower. It may also be that the mean $v_{\rm{enc}}$ is greater than the predicted $v_{\rm{enc}}$ because of the gravitational acceleration, which is ignored in the analytical model, so that $\eta$ is smaller than its prediction. Because the pole/equator asymmetry is not seen in the $V$ distribution and substituting the predicted $v_{\rm{enc}}$ with the simulated $\bar{V}$ only reduces the reproduction error to about 27\%, other places where the numerical simulations deviate from the analytical model need to be checked. Although statistical fluctuation results in a great error of $\eta$ reproduced from $A_1^{\Theta}$, fortunately the reproductions from asymmetry amplitudes of $V_{\bot}$, $D$, $N$, and $N_{\rm{c}}$ are much better, especially for $N$, which leads to reproduction errors of 1\%--15\% for all cases.

Again we followed the method of reproducing bombardment conditions described in Sect. \ref{subsec-rep} and present a simple estimation based on the fit $A_{0,1}^{D,\rm{c}}$ of case 1, for only $D$ and $N_{\rm{c}}$ are observables in surveys of crater records. Assuming $v_{\rm{enc}} = 8.32 $ km~s$^{-1}$ and $t = 10^{7}$ yr are known, the reproduced conditions are calculated to be $v_{\rm{M}} = 1.45 \pm 0.02$ km~s$^{-1}$, $\alpha_{\rm{p}} = 3.23 \pm 0.06$, $d_{\rm{min}} = 4.71 \pm 0.03$ km, and $F = (1.821 \pm 0.164) \times 10^{-4}$ yr$^{-1}$. Comparisons of the former three to the predictions, $v_{\rm{M}}$ = 1.78 km~s$^{-1}$, $\alpha_{\rm{p}} = 3$, and $d_{\rm{min}}$ = 5 km, lead to differences of 19\%, 8\%, and 6\%, respectively. We note that the impactor flux $F$ is derived with Eq. \ref{eq-A0c} where $l_{\rm{gl}} = 2 \pi R_{\rm{M}}$ is substituted by $S_{\rm{gl}} = 4 \pi R_{\rm{M}}^2$, as the numerical simulations are three-dimensional. It is equivalent to a total number of impacts of $1821 \pm 164$ during an integration time of $10^{7}$ yr, which is very close to the actual result $C = 1544$. Taking the limited number of impacts in simulations into account, it is rather inspiring to find that this method of reproducing the bombardment conditions from the cratering distribution works well. The importance of this method is fully shown when we recall that very many lunar craters were generated by MBAs during the LHB, meaning that our current knowledge of the LHB and MBAs may provide clues to the early lunar dynamical history.

%

\subsection{Coupling of leading/trailing and pole/equator asymmetries} \label{subsec-couple}

\begin{figure*}
    \centering
    \includegraphics[width=17cm]{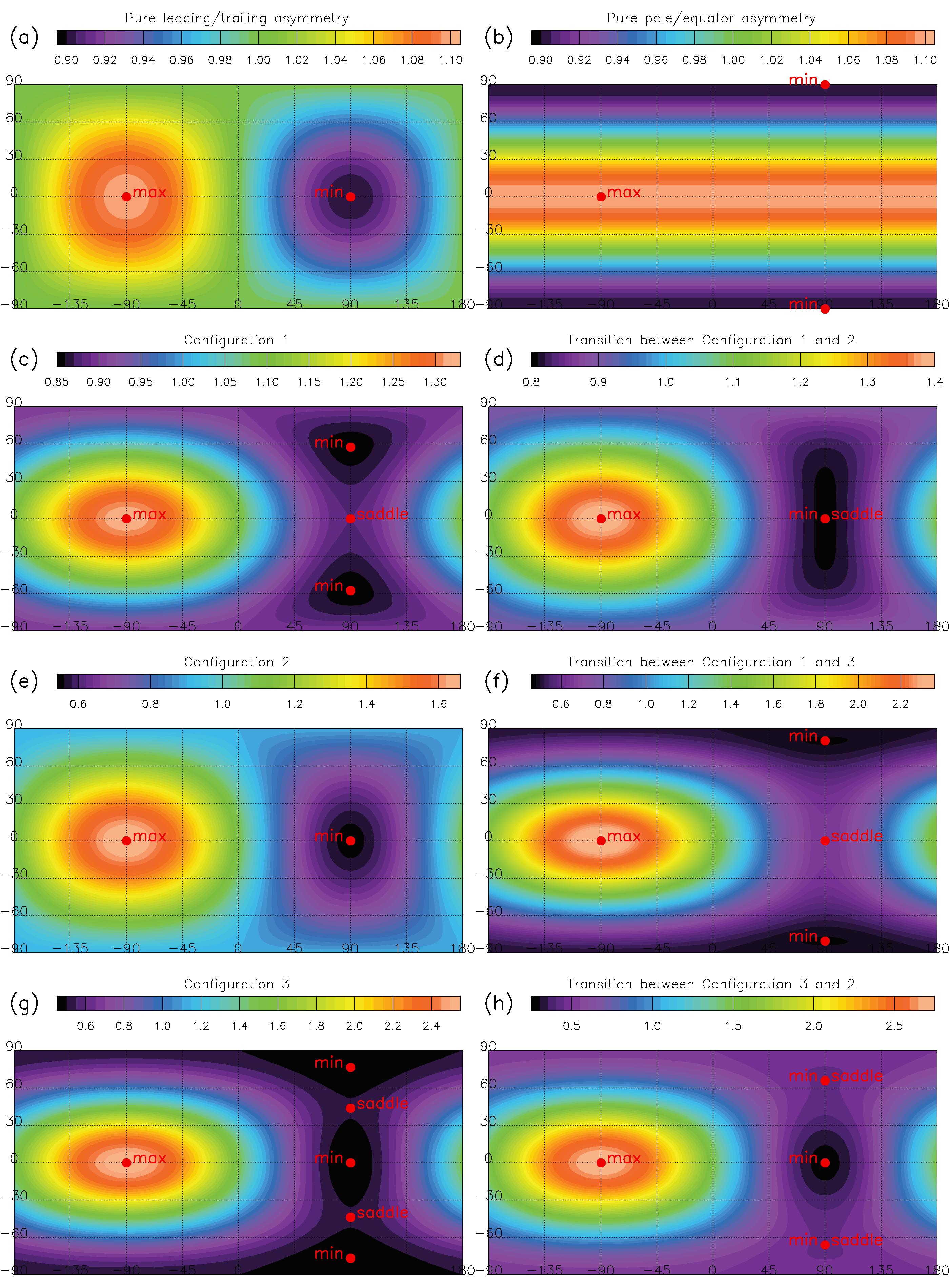}
    \caption{Normalized cratering distribution of the coupled asymmetries.
    a) Pure leading/trailing asymmetry with $A_2 = 0$;
    b) pure pole/equator asymmetry with $A_1 = 0$;
    c) configuration 1 with $A_2 \le \frac{1}{7}$ and $A_1 \le \frac{4A_2}{5A_2+1}$;
    d) transition between configurations 1 and 2 with $A_2 \le \frac{1}{7}$ and $A_1 = \frac{4A_2}{5A_2+1}$;
    e) configuration 2 with $A_2 \le \frac{1}{7}$ and $A_1 > \frac{4A_2}{5A_2+1}$;
    f) transition between configurations 1 and 3 with $A_2 \in (\frac{1}{7},\frac{3}{5}]$ and $A_1 = \frac{4A_2}{5A_2+1}$;
    g) configuration 3 with $A_2 \in (\frac{1}{7},\frac{3}{5}]$ and $A_1 \in (\frac{4A_2}{5A_2+1}, \sqrt{\frac{2A_2}{3(1-A_2)}}]$;
    h) transition between configurations 3 and 2 with $A_2 \in (\frac{1}{7},\frac{3}{5}]$ and $A_1 = \sqrt{\frac{2A_2}{3(1-A_2)}}$.}
    \label{fig-couple}
\end{figure*}

Fig. \ref{fig-Gamma sim} illustrates the spatial distributions of $\Gamma$. All of the distributions involve the leading/trailing asymmetry, and all but the $V$ distribution also show the pole/equator asymmetry. The coupling of the two asymmetries can be described using Eq. \ref{eq-couple}, whose normalized form is
\begin{equation}    \label{eq-couple /A0}
\begin{split}
  \Delta\Gamma(\lambda,\varphi) & = (1 + A_1 \cos\beta) (1 + A_2 \cos2\varphi) \\
    & = (1 - A_1 \sin\lambda \cos\varphi) (1 + A_2 \cos2\varphi)
\end{split}
\end{equation}
The asymmetry amplitudes $A_{1,2}$ must be in $[0,1]$ to ensure that $\Delta\Gamma$ is non-negative and its first and second factors are in negative correlations with $\beta$ and $|\varphi|$. Clearly, the coupled distribution is symmetric about both the equator and the apex-antapex axis, with a maximum $(1 + A_1)(1 + A_2)$ always at the apex. If $\varphi$ is constant, the function degenerates to $\Delta\Gamma = A_0' (1 - A_1' \sin\lambda)$, where $A_0' = 1 + A_2 \cos2\varphi$ and $A_1' = A_1 \cos\varphi$. This means that on every latitude line, $\Delta\Gamma$ is maximized and minimized at longitude $-90\degr$ and $+90\degr$ with the average $A_0'$ and the variation amplitude $A_1'$ dependent on $|\varphi|$. If $\lambda$ is constant, the function is a cubic polynomial of $\cos\varphi$, whose first factor monotonically decreases or increases as $|\varphi|$ increases on the leading or trailing side, respectively, and the second factor always decreases with increasing $|\varphi|$. Therefore, with a given $\lambda \in (-180\degr, 0\degr)$, $\Delta\Gamma$ is always maximized on the equator and minimized on the poles, while with a given $\lambda \in (0\degr, 180\degr)$, the conditions vary with varying $A_{1,2}$. The distributions are thus classified into three configurations according to the numbers and locations of the extrema.

We summarize the theory distribution of the coupled asymmetries as follows. The case $A_1 \neq 0$ and $A_2 = 0$ corresponds to the pure leading/trailing asymmetry, when the contour lines on the lunar surface are all circles centered on the apex and the antapex, which are the locations of the only maximum and minimum (Fig. \ref{fig-couple}a). The case $A_1 = 0$ and $A_2 \neq 0$ corresponds to the pure pole/equator asymmetry, leaving a distribution consisting of latitude-parallel strips with a peak at the whole equator and two minima at the poles (Fig. \ref{fig-couple}b). Excluding those two limiting conditions, there are two critical values of $A_2$, $\frac{1}{7}$ and $\frac{3}{5}$, and two critical values of $A_1$ with $A_2$ given, $\frac{4A_2}{5A_2+1}$ and $\sqrt{\frac{2A_2}{3(1-A_2)}}$, which together define the configurations. We note that $\frac{4A_2}{5A_2+1} \le \sqrt{\frac{2A_2}{3(1-A_2)}}$ always holds and $\frac{4A_2}{5A_2+1} = \sqrt{\frac{2A_2}{3(1-A_2)}} = \frac{1}{3}$ only when $A_2=\frac{1}{7}$.
\begin{enumerate}
  \item When $A_2 \in (0,\frac{1}{7}]$ is fixed, two configurations appear in turn as $A_1$ increases. When $A_1 \in (0, \frac{4A_2}{5A_2+1}]$, the contours begin twisting from the strips (Fig. \ref{fig-couple}c). The maximum can only be reached at the apex $(-90 \degr, 0 \degr)$, while the minima slide symmetrically along the longitude $90\degr$ from the poles toward the antapex $(90 \degr, 0 \degr)$. The latitudes of the two minima are determined by
      \begin{equation}  \label{eq-phi_min}
        \cos\varphi = \frac{1}{3A_1} - \sqrt{\frac{2A_2 - 3(1-A_2)A_1^2}{18A_1^2A_2}}.
      \end{equation}
      The antapex acts as a saddle point. We call this distribution configuration 1. The moment $A_1$ increases to $\frac{4A_2}{5A_2+1}$ is when the minima join at the antapex and become one (Fig. \ref{fig-couple}d). After that when $A_1 \in (\frac{4A_2}{5A_2+1}, 1]$, there is only one maximum $(1+A_1)(1+A_2)$ at the apex and one minimum $(1-A_1)(1+A_2)$ at the antapex (Fig. \ref{fig-couple}e). The contours near the apex and antapex resemble lying and standing ellipses. The higher the value of $A_1$, the closer to circles the contours. The distribution at this stage is called configuration 2.
  \item When $A_2 \in (\frac{1}{7},\frac{3}{5}]$ is fixed, another configuration presents itself as well as the above two. When $A_1 \in (0, \frac{4A_2}{5A_2+1}]$, configuration 1 is also observed, even though at the moment $A_1=\frac{4A_2}{5A_2+1}$ the two minima are still approaching each other. At this moment, the saddle at the antapex becomes two saddles that start approaching the minima, and the antapex itself becomes a third minimum (Fig. \ref{fig-couple}f). As $A_1 \in (\frac{4A_2}{5A_2+1}, \sqrt{\frac{2A_2}{3(1-A_2)}}]$ increases, the saddle and minimum on each side of the equator come increasingly closer, while the antapex valley continuously deepens. This stage corresponds to configuration 3 (Fig. \ref{fig-couple}g), when the latitudes of the saddles are determined by
      \begin{equation}
        \cos\varphi = \frac{1}{3A_1} + \sqrt{\frac{2A_2 - 3(1-A_2)A_1^2}{18A_1^2A_2}}.
      \end{equation}
      The moment $A_1 = \sqrt{\frac{2A_2}{3(1-A_2)}}$ is when each pair of approaching saddle and minimum combines and vanishes at the points $(90\degr, \pm \arccos(\frac{1}{3A_1}))$ (Fig. \ref{fig-couple}h). When $A_1 \in (\sqrt{\frac{2A_2}{3(1-A_2)}}, 1]$, the distribution is that of configuration 2 based on our simple classification. However, it is necessary to point out that near the vanishing points, the contour line shapes are different from those with $A_2 \in (0, \frac{1}{7}]$.
  \item When $A_2 \in (\frac{3}{5},1]$ is fixed, the last stage of case $A_2 \in (\frac{1}{7},\frac{3}{5}]$ belonging to configuration 2 will not appear. The reason is that only if $A_2 \le \frac{3}{5}$ can the critical value $\sqrt{\frac{2A_2}{3(1-A_2)}} \le 1$. When $A_2 \in (\frac{3}{5},1]$, $A_1$ can never reach $\sqrt{\frac{2A_2}{3(1-A_2)}}$ before increasing to 1.
\end{enumerate}

Simply speaking, configurations 1, 2, and 3 are defined to have two, one, and three minima each and one maximum always. Our fit distributions of $\Theta$, $V_\bot$, $D$, and $N$ all belong to configuration 1 and 2 since none of their $A_2$ is larger than $\frac{1}{7}$ in any simulation case. In particular, the $\Theta$ and $D$ distributions of all the cases belong to configuration 1. The $N_{\rm{c}}$ distribution, with $A_2^{\rm{c}}$ always larger than $\frac{1}{7}$, belongs to configuration 2 for cases 1--3, configuration 1 for case 4 and configuration 3 for case 5. We note that we classify those distributions according to their fit $A_1$ and $A_2$, but some simulated distributions are not quite symmetric about the equator, and therefore the pole/equator asymmetry amplitudes are sometimes underestimated by their fits. Nearly all the simulated distributions clearly exhibit the main properties of configuration 1 or 2 and are well fit. However, we cannot visually verify whether the $N_{\rm{c}}$ distribution of case 5 has a third minimum at the antapex, resulting form the small number statistics. More detailed studies in terms of both theory and observation are necessary to check the description of coupled-asymmetric distribution.

\section{Discussion}    \label{sec-discu}

\subsection{Comparison between NEOs and MBAs}   \label{subsec-NEOs vs MBAs}

\begin{figure}
    \resizebox{\hsize}{!}{\includegraphics{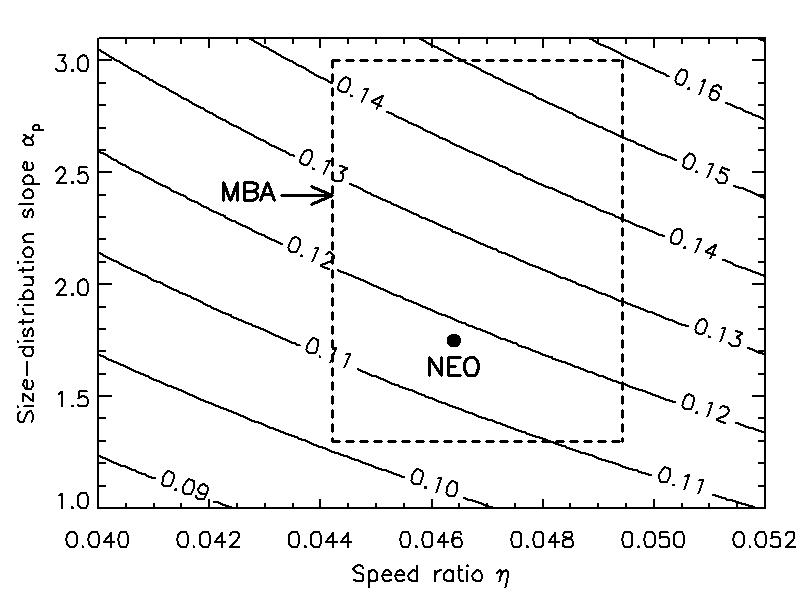}}
    \caption{Contour map of the leading/trailing asymmetry amplitude of the crater density $A_1^{\rm{c}}$ as a function of speed ratio $\eta$ and size-distribution slope $\alpha_{\rm{p}}$. The estimates of $A_1^{\rm{c}}$ for NEOs and MBAs are indicated by dotted and dashed squares, respectively.}
    \label{fig-A1c}
\end{figure}

\citet{Strom2005, Strom2015} suggested that the MBAs and NEOs that dominated during the LHB and since about 3.8--3.7 Gya in turn had formed craters on the Moon and even all the terrestrial planets. We review the current knowledge of two impactor populations in terms of orbital distribution, size distribution, etc. in Sect. \ref{sec-intro}. Now that we have formulated the cratering asymmetry based on our analytical model in Sect. \ref{sec-analy}, which has been verified in various aspects by numerical simulations in Sect. \ref{sec-simul}, we analytically estimate and compare the leading/trailing asymmetry amplitudes generated by the two impactor populations in this subsection.

Given a population of impactors, the typical encounter speed $v_{\rm{enc}}$ is determined by its orbital distribution, the minimum size $d_{\rm{min}}$ and slope $\alpha_{\rm{p}}$ are determined by its size distribution, and the flux $F$ is determined by both. The first three parameters influence the spatial distribution of $D$, and all of the four are involved in the $N_{\rm{c}}$ distribution. When the normalized distribution depending on just $A_1$ is considered, then only $v_{\rm{enc}}$ and $\alpha_{\rm{p}}$ matter.

Applying our analytical model, the inclination is not considered, and because we lack complete information about the orbital distribution of MBAs 3.9 Gya, we only take the difference between MBAs and NEOs in semi-major axis $a_{\rm{p}}$ into account, that is, the distributions of perihelion distance $q_{\rm{p}}$ are assumed to be the same. The semi-major axes of MBAs lie in the range of 2.0--3.5 AU, while those of NEOs lie in the range of 0.5--2.8 AU according to the "constrained target region" of the debiased NEO model built by \citet{Bottke2002}. When we adopt the medians of $a_{\rm{p}}$, 2.75 AU for MBAs and 1.65 AU for NEOs, and a median of $q_{\rm{p}}$, 0.5 AU, the typical encounter speeds of MBAs and NEOs are calculated to be 25.4 km~s$^{-1}$ and 22.1 km~s$^{-1}$ (Eq. \ref{eq-v_enc}), and thus their typical impact speeds are 25.5 km~s$^{-1}$ and 22.3 km~s$^{-1}$ ($v_{\rm{imp}}^2 = v_{\rm{enc}}^2 + v_{\rm{esc}}^2$). This estimate of $v_{\rm{imp}}$ for NEOs is consistent with results of \citet{Gallant2009}, \citet{Ito2010} and \citet{LeFeuvre2011}, who found the mean impact speed to be 20 km~s$^{-1}$, 22.4 km~s$^{-1}$, and 19.7 km~s$^{-1}$, respectively.
The lunar orbital speeds $v_{\rm{M}}$ during the dominant epochs of the two impactor populations are different as well. For MBAs, $v_{\rm{M}}$ during the LHB may be from 1.12 to 1.26 km~s$^{-1}$, relevant to the uncertain $a_{\rm{M}}$ from 50 to 40 $R_{\oplus}$ (Sect. \ref{sec-intro}). For NEOs, the current value $v_{\rm{M}}$ = 1.03 km~s$^{-1}$ is used since $a_{\rm{M}}$ is believed to have been beyond 40 $R_{\oplus}$ for most of the Moon lifetime. Here $e_{\rm{M}}$ is ignored in the calculation.

Therefore, the speed ratio $\eta$ is 0.044--0.049 for MBAs and 0.046 for NEOs. A clear excess or deficiency of $\eta$ of MBAs compared to NEOs is not seen. This is also true for the asymmetry amplitude $A_1^D$ because it is proportional to $\eta$ (Eq. \ref{eq-A1D}): $A_1^D$ is 0.0205--0.0229 for MBAs and 0.0215 for NEOs. Conversely, the fit $A_1^D$ of an observed area on the lunar surface that is purely cratered by MBAs can lead to a speed ratio and thus an Earth-Moon distance during the LHB, helping to clarify the early lunar history. On the other hand, when $A_1^D$ of MBAs is precisely determined in theory, the fit $A_1^D$ of a cratered area can lead to the realistic ratio of crater numbers of the two impactor populations.

The asymmetry amplitude $A_1^{\rm{c}}$ depends on not only $\eta$, but also on $\alpha_{\rm{p}}$ (Eq. \ref{eq-A1c}). Assuming the size distribution slope of the NEOs is $\alpha_{\rm{p}} = 1.75$ \citep{Bottke2002}, Eq. \ref{eq-slope p2c} leads to the slope of craters generated by them of $\alpha_{\rm{c}}$ = 2.07, which is well consistent with \citet{Strom2005,Strom2015}, who found that the size distribution of craters on lunar young plains has a single slope of 2. Thus the $A_1^{\rm{c}}$ of NEOs is calculated to be 0.117. The size distribution of the MBAs is not a single power law as mentioned in Sect. \ref{sec-intro}, and the variation of its slope is not slight, as \citet{Ivezic2001} suggested for $d_{\rm{p}}$ = 0.4--5 km and 5--40 km, $\alpha_{\rm{p}}$ = 1.3 and 3, respectively. Still, all of our analysis is valid for the separated part of impactors in a given $d_{\rm{p}}$ interval with invariant $\alpha_{\rm{p}}$ and the craters they generate. Therefore, the possible range of $A_1^{\rm{c}}$ of MBAs shown in Fig. \ref{fig-A1c} is defined by the top-right maximum 0.159, where $\eta$ = 0.049 ($a_{\rm{M}}$ = 40 $R_{\oplus}$) and $\alpha_{\rm{p}} = 3$, and the bottom left minimum 0.101, where $\eta$ = 0.044 ($a_{\rm{M}}$ = 50 $R_{\oplus}$) and $\alpha_{\rm{p}} = 1.3$. For the considered part of MBAs with $d_{\rm{p}}$ = 0.4--40 km, Fig. \ref{fig-A1c} shows that their $A_1^{\rm{c}}$ is probably greater than that of the NEOs. The $N_{\rm{c}}$ distribution generated by impactors with a size distribution of multiple power laws will be formulated in a future work.

We caution that even the size distribution is not a single power law, and thus $\bar{d}_{\rm{p}} \neq \frac{\alpha_{\rm{p}}}{\alpha_{\rm{p}}-1}d_{\rm{min}}$, the definition $\bar{d}_{\rm{p}} = \int_{+\infty}^{d_{\rm{min}}} d_{\rm{p}} {\rm{d}} \bar{N}_{\rm{p}}$ (Eq. \ref{eq-d_p mean}) still holds, so that the formulations of the $D$ distribution and $\bar{D}$ also apply. The approximate $\bar{D}$ (Eq. \ref{eq-D glo appr}) only depends on $\bar{d}_{\rm{p}}$ and $v_{\rm{enc}}$. Because $v_{\rm{enc}}$ of the MBAs is greater than that of NEOs and MBAs are generally larger than NEOs, this results in the separation into two crater populations in terms of size. A rough estimation leads to $\bar{D} \sim$ 33 km for MBAs and $\bar{D} \sim$ 4 km for NEOs, assuming $\bar{d}_{\rm{p}} \sim$ 1 km and $\bar{d}_{\rm{p}} \sim$ 0.1 km, respectively.

\subsection{Apex/antapex and pole/equator ratios}

\begin{figure}
    \resizebox{\hsize}{!}{\includegraphics{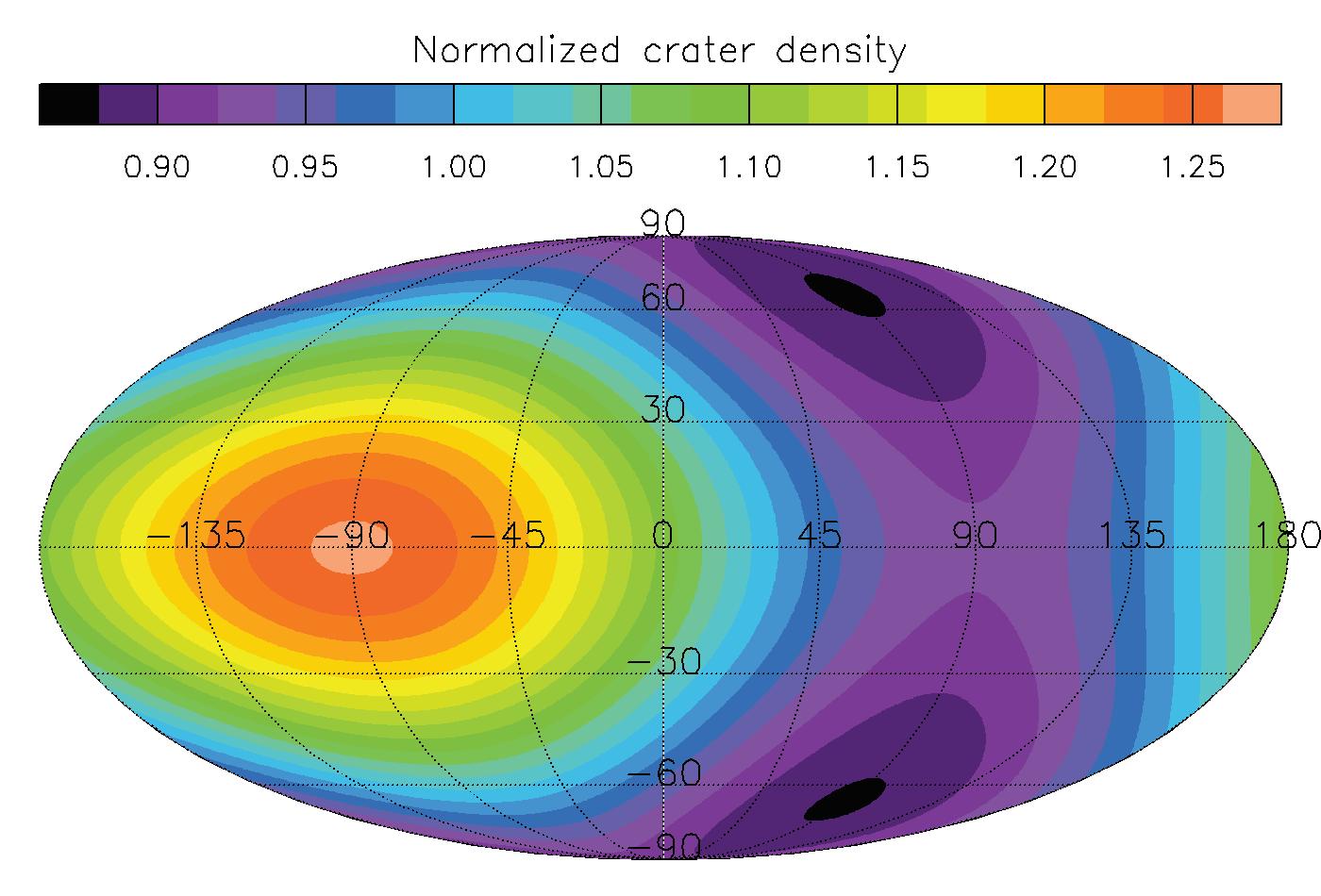}}
    \caption{Reproduction of spatial distribution of the crater density in \citet{LeFeuvre2011}. It is generated using Eq. \ref{eq-couple /A0} with asymmetry amplitudes $A_1$ = 0.156 and $A_2$ = 0.093.}
    \label{fig-LF11}
\end{figure}

\begin{figure}
    \resizebox{\hsize}{!}{\includegraphics{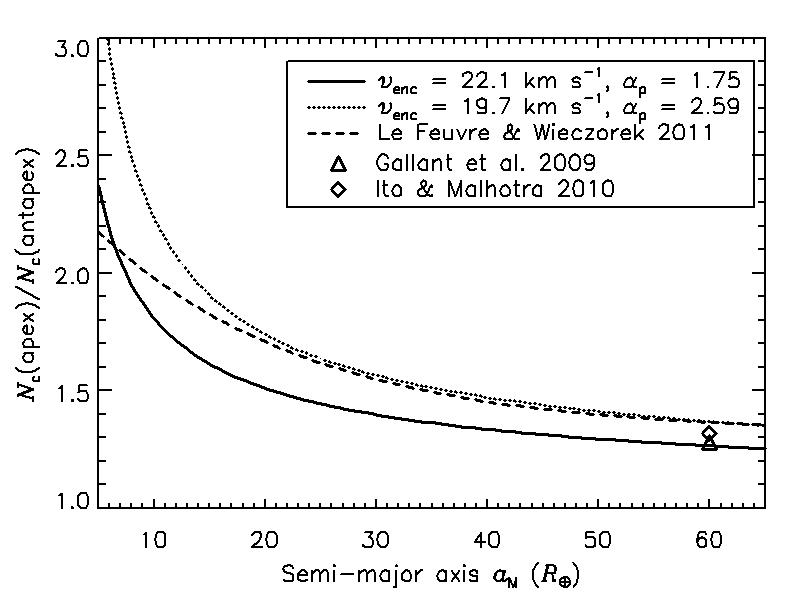}}
    \caption{Apex/antapex ratio of the crater density generated by the current impactors in related works compared with ours. Our relation between $N_{\rm{c}}({\rm{apex}})/N_{\rm{c}}({\rm{antapex}})$ and $a_{\rm{M}}$ with $v_{\rm{enc}} = 22.1$ km~s$^{-1}$ and $\alpha_{\rm{p}} = 1.75$ (solid curve) is close to the empirical one in \citet{LeFeuvre2011}, which works for $a_{\rm{M}}$ = 20--60 $R_{\oplus}$ (Eq. \ref{eq-LF11}; dashed curve); the former with $v_{\rm{enc}} = 19.7$ km~s$^{-1}$ and $\alpha_{\rm{p}} = 2.59$ (dotted curve) is even nearly equal to the latter in its working range. Simulation results from \citet{Gallant2009} and \citet{Ito2010} assuming the current lunar orbit (triangle and square) are also indicated.}
    \label{fig-Nc ratio}
\end{figure}

Earlier studies on cratering have often used the apex/antapex and pole/equator ratios to indicate the asymmetry degree of the spatial distribution. The former is the ratio of the value (typically the crater density $N_{\rm{c}}$) at the apex to the antapex, which increases with increasing leading/trailing asymmetry, while the latter is the ratio of the value at the poles to the equator, which increases with decreasing pole/equator asymmetry. These ratios can be obtained easily and are especially convenient for simulations and observations, but their physical meanings were not understood before. Neither was it possible to completely define a spatial distribution through them. According to the formulation of coupled asymmetries that we established based on our analytical model and simulation results (Eq. \ref{eq-couple}), the ratios are each related to the relevant asymmetry amplitude by
\begin{align}
  \Gamma({\rm{apex}})/\Gamma({\rm{antapex}}) &= \frac{1 + A_1}{1 - A_1},   \label{eq-a/a ratio}\\
  \Gamma({\rm{pole}})/\Gamma({\rm{equator}}) &= \frac{1 - A_2}{1 + A_2}.
\end{align}
(When $A_1 \neq 0$, $\Gamma({\rm{equator}})$ means the average over the equator.) Because a whole distribution can be directly reproduced given $A_1$ and $A_2$ using Eq. \ref{eq-couple /A0} and because the meaning of $A_1$ has explicitly been determined in Sect. \ref{subsec-appr}, the asymmetry amplitudes are the better measurements of the asymmetry degrees for theories.

We now reproduce the spatial distribution of $N_{\rm{c}}$ shown in Fig. 5 of \citet{LeFeuvre2011} through asymmetry amplitudes. They obtained it by assuming the current Earth-Moon distance and the current asteroids and comets in the inner solar system as impactors. The value of $N_{\rm{c}}$ varies from 0.80 at $(90\degr, \pm65\degr)$ to 1.25 at the apex in terms of the global average, which leads to a maximum-minimum ratio of 1.5, an apex/antapex ratio of 1.37, and a pole/equator ratio of 0.80. Using their apex/antapex ratio and minimum location, we reproduce the distribution shown in Fig. \ref{fig-LF11}. It is derived from Eq. \ref{eq-couple /A0} with the asymmetry amplitudes $A_1^{\rm{c}}$ = 0.156 and $A_2^{\rm{c}}$ = 0.093, which are calculated with Eqs. \ref{eq-a/a ratio} and \ref{eq-phi_min} to ensure $N_{\rm{c}}({\rm{apex}})/N_{\rm{c}}({\rm{antapex}})$ = 1.37 and the latitude of the minimum $\varphi_{\rm{min}} = \pm 65\degr$. Then its maximum and minima are 1.26 at apex and 0.88 at $(90\degr, \pm65\degr)$, whose ratio is 1.44, well consistent with that of \citet{LeFeuvre2011}, and our pole/equator ratio 0.83 is nearly equal to theirs. Not only the $N_{\rm{c}}$ distribution is well reproduced, but also the estimate $\alpha_{\rm{p}}$ = 2.6 is derived from Eq. \ref{eq-A1c} with the reproduced $A_1^{\rm{c}}$, current $v_{\rm{M}}$, and the mean impact speed on the Moon $v_{\rm{enc}}$ = 19.7 km~s$^{-1}$ according to \citet{LeFeuvre2011}. Therefore, a power-law size distribution with this slope is found to be a convenient substitute for the 10th-order polynomial that \citet{LeFeuvre2011} used for the empirical fit in their approach.

We have determined the leading/trailing asymmetry amplitude of the $N_{\rm{c}}$ distribution generated by NEOs to be $A_1^{\rm{c}}$ = 0.117 (Sect. \ref{subsec-NEOs vs MBAs}), which is equivalent to $N_{\rm{c}}({\rm{apex}}) /N_{\rm{c}}({\rm{antapex}})$ = 1.27. As reviewed in Sect. \ref{sec-intro}, \citet{Morota2003} reported the apex/antapex ratio to be $\sim1.5$ based on observations of young craters on the lunar highlands, \citet{Gallant2009} and \citet{Ito2010} found the ratio to be $1.28\pm0.01$ and $1.32\pm0.01$ from their numerical simulations, and \citet{LeFeuvre2011} calculated it to be 1.37 using their semi-analytical approach, all in good consistence with our estimate. Moreover, \citet{LeFeuvre2011} also empirically fit the dependence of $N_{\rm{c}}({\rm{apex}})/N_{\rm{c}}({\rm{antapex}})$ on $a_{\rm{M}}$ and obtained Eq. \ref{eq-LF11} that is valid for $a_{\rm{M}}$ = 20--60 $R_{\oplus}$. Treating the expression of $A_1^{\rm{c}}$ (Eq. \ref{eq-A1c}) as a function of $a_{\rm{M}}$, that is, $A_1^{\rm{c}}(a_{\rm{M}}) = A_1^{\rm{c}}(v_{\rm{M}}(a_{\rm{M}}))$ where $v_{\rm{M}}(a_{\rm{M}}) = \sqrt{G (m_{\oplus}+m_{\rm{M}}) /a_{\rm{M}}}$, we can also derive a relation
\begin{equation}
    N_{\rm{c}}({\rm{apex}})/N_{\rm{c}}({\rm{antapex}}) = \frac{1 + A_1^{\rm{c}}(a_{\rm{M}})}{1 - A_1^{\rm{c}}(a_{\rm{M}})}.
\end{equation}
The two relations are compared in Fig. \ref{fig-Nc ratio} with solid and dashed lines. We note that in calculating our relation, $v_{\rm{enc}}$ = 22.1 km~s$^{-1}$ and $\alpha_{\rm{p}}$ = 1.75 are the same as used in Sect. \ref{subsec-NEOs vs MBAs}, but $a_{\rm{M}}$ is no longer fixed at 60 $R_{\oplus}$ since it is now considered a variable. Figure \ref{fig-Nc ratio} shows that the two relations are quite close especially where that of \citet{LeFeuvre2011} is valid. When we substitute their mean impact speed on the Moon of 19.7 km~s$^{-1}$ for our roughly estimated $v_{\rm{enc}}$ and adopt $\alpha_{\rm{p}}$ = 2.6 as reproduced from their $N_{\rm{c}}$ distribution, our function is identical to their valid part with a difference of only $<$ 2\%.

We briefly point out the difference between our work and that of \citet{LeFeuvre2011} in cratering modeling. Their model is able to produce the pole/equator asymmetry since the anisotropic impactor population was adopted, while our current analytical model alone can only formulate the leading/trailing asymmetry, but with our empirical description of the pole/equator asymmetry based on numerical simulations, we can also describe, reproduce, and measure the coupled asymmetries. More importantly, they integrated the cratering distribution numerically, while we give explicit analytical functions after rigorous deduction, which clearly shows the influence of every factor and makes the formulations very easy to apply.

\subsection{Generalization of the analytical model}

In addition to the Earth-Moon system, our analytical model (Sect. \ref{sec-analy}) can be applied to other planet-satellite systems. In the inner solar system, the NEOs and MBAs are the common impactor populations dominating different epochs. They encounter Mars with $v_{\rm{enc}} \approx$ 19.3 and 22.9 km~s$^{-1}$, respectively, which is calculated with $a_{\oplus}$ replaced by the semi-major axis of Mars $a_{\rm{Mars}}$ = 1.5 AU in Eq. \ref{eq-v_enc} and the typical $a_{\rm{p}}$ of NEOs being the median of the interval 0.76--2.8 AU instead of 0.5--2.8 AU because projectiles with $a_{\rm{p}} < a_{\rm{Mars}}/2$ on elliptic orbits cannot encounter Mars. The Martian moons, Phobos and Deimos, which lie on very circular orbits, have current orbital speeds of $v_{\rm{orb}}$ = 2.14 and 1.35 km~s$^{-1}$ and negligible escape speeds $v_{\rm{esc}}$. Thus, the NEOs with a typical speed ratio $\eta$ of 0.111 for Phobos and 0.070 for Deimos are expected to generate a current leading/trailing asymmetry amplitude of the crater density of 0.28 and 0.18, respectively, both of which are much greater than for the Moon. Similarly, the orbital speeds of Phobos and Deimos and even their orbits during the LHB might be derived from the observed crater records if they had been tidal-locked then.

The dominant impactors in the outer solar system are ecliptic comets, which are generally considered to originate in the Kuiper belt \citep{Zahnle2001}. Because their encounter speeds are generally lower than the orbital speeds of the satellites of the giants, our analytical model is not completely applicable. The divergence starts from the behavior of the normal impact point. With our assumption $v_{\rm{enc}} > v_{\rm{orb}}$, the normal impact point moving westward along the satellite's equator (coplanar with its orbit and ecliptic) reaches the antapex and apex when the satellite is at its pericenter (where $\vec{v}_{\rm{enc}}$ is in the same direction as $\vec{v}_{\rm{orb}}$) and apocenter, while the impact speed $v$ is minimized and maximized (Sect. \ref{subsec-imp}). However, if $v_{\rm{enc}} < v_{\rm{orb}}$, the normal impact point reaches the apex at both the pericenter and apocenter. Specifically, as the satellite's true anomaly $f$ increases from 0$\degr$ (pericenter), the normal impact point leaves the apex, moves eastward until $f = \arccos (v_{\rm{enc}}/v_{\rm{orb}}) \in (0\degr, 90\degr)$, turns around and returns to the apex when $f = 180\degr$; as $f$ continues increasing, it continues moving westward until $f = 360\degr - \arccos (v_{\rm{enc}}/v_{\rm{orb}}) \in (270\degr, 360\degr)$ and then returns to the apex when $f = 360\degr$. Its path is symmetric about the apex, bounded by the two return points at equal distances of $\beta_{\rm{max}} = \arctan (v_{\rm{enc}}/\sqrt{v_{\rm{orb}}^2 - v_{\rm{enc}}^2})$ from it. It is seen that $\beta_{\rm{max}} < 90\degr$, so the satellite surface is divided into four areas: the area centered on the antapex with the longitude $\lambda \in (\beta_{\rm{max}}, 180\degr - \beta_{\rm{max}})$ never is hit by impactors; the area $(\beta_{\rm{max}} - 180\degr, -\beta_{\rm{max}})$ centered on the apex is always under bombardment; and the two remaining symmetric areas on the near and far side each are periodically included in and excluded from the instant bombarded hemisphere. Apparently, the higher the ratio $v_{\rm{orb}}/v_{\rm{enc}}$, the smaller the distance $\beta_{\rm{max}}$, and thus the greater the asymmetry. As a result, the leading/trailing asymmetry is much enhanced when $v_{\rm{enc}} < v_{\rm{orb}}$ than otherwise. A generalized model including this case will be established in the future to examine the cratering of giants' satellites.

\citet{Zahnle2001} suggested a semi-empirical description of $N_{\rm{c}}$ distribution on satellites of giant planets shown as Eq. \ref{eq-Zahnle2001}. We find qualitative consistence of our approximate formulation $N_{\rm{c}} (\beta)$ (Eq. \ref{eq-N_c appr}) with theirs since the asymmetry is amplified by speed ratio $v_{\rm{orb}}/v_{\rm{enc}}$ and size-distribution slope $\alpha_{\rm{p}}$ in both forms. One difference is that $\alpha_{\rm{p}}$ is a coefficient of $\cos\beta$ in our form but acts as an exponent in theirs. Considering they derived their formulation based on simulation results, and referring to our analysis, there is a great possibility that $\alpha_{\rm{p}}$ is still a coefficient (in approximate description) when $v_{\rm{enc}} < v_{\rm{orb}}$ and the influence of $\alpha_{\rm{p}}$ was overestimated by formulation of \citet{Zahnle2001}, which will be examined by future model.

\section{Conclusions}

We analytically formulated the lunar cratering distribution and confirmed the derivation with numerical simulations. The formulations are quite easy to use. They are able to give results nearly identical to related works, avoiding the time-consuming simulations and clarifying the physical meanings in quantitative relations.

Based on a planar model excluding the gravitations of Earth and the Moon on the impactors, we derived series of formulations of the cratering distribution on the synchronously rotating satellite through rigorous integration. The formulations directly and unambiguously proved the existence of a leading/trailing cratering asymmetry and the identity of the near and far sides in all aspects of cratering, that is, impact speed $V$, incidence angle $\Theta$, normal speed $V_{\bot}$, crater diameter $D$, impact density $N$, and crater density $N_{\rm{c}}$. Series expansion to the first order of $v_{\rm{M}}/v_{\rm{enc}}$ resulted in the approximate $\Gamma$ distributions in the common form $\Gamma (\beta) = \bar{\Gamma} (1 + A_1 \cos\beta)$, where $\bar{\Gamma}$ is in positive correlation with $v_{\rm{enc}}$, except that $\bar{\Theta}$ is the constant 51.8$\degr$, and $A_1 \propto v_{\rm{M}}/v_{\rm{enc}}$, that is, $A_1 \propto a_{\rm{M}}^{-1/2}$ with a given $v_{\rm{enc}}$. The relations between the cratering distribution and the bombardment conditions including properties of impactor population and the lunar orbit during the bombardment provides a viable method of reproducing the latter by measuring the former. The lower the speed ratio $v_{\rm{M}}/v_{\rm{enc}}$, the better the approximation to the exact $\Gamma$ distributions and the better the reproduction of the bombardment conditions. In particular, the approximation of $N_{\rm{c}}$ can be improved by decreasing $\alpha_{\rm{p}}$ as well.

We also numerically simulated the cratering on the Moon that is caused by the MBAs during the LHB with five simulation cases in which $a_{\rm{M}}$ = 20--60 $R_{\oplus}$. Not only the leading/trailing asymmetry is present in the simulated cratering distribution in all aspects, but also the pole/equator asymmetry is seen, except for the $V$ distribution, while the signs of the near/far asymmetry are not enough. The pole/equator asymmetry is empirically formulated with $(1 + A_2 \cos2\varphi)$, leading to the description of coupled asymmetries $\Gamma(\lambda,\varphi) = \bar{\Gamma} (1 + A_1 \cos\beta)(1 + A_2 \cos2\varphi)$, which is well fit to the simulated $\Gamma$ distributions with errors of $\bar{\Gamma}$ smaller than 6\%. For each case, the analytical prediction of $\bar{V}$ is only a few percent lower than the simulated $\bar{V}$, but the predicted $\bar{\Theta}$ is obviously larger than the simulated $\bar{\Theta}$ because the pole/equator asymmetry is involved. As predicted by the analytical model, $\bar{\Gamma}$ except for $\bar{\Theta}$ all show negative correlations with $a_{\rm{M}}$ in general, and so do the fit $A_1$ except for $A_1^{\Theta}$, whose variation with $a_{\rm{M}}$ involves greater statistical fluctuations, while the fit $A_2$ seem independent of $a_{\rm{M}}$. Additionally, the method of reproducing performs well, with reproduction errors being $\sim$10\% for bombardment conditions in case 1. This means that it is possible to speculate about the lunar orbital status during the LHB when the cratering distribution generated by MBAs is determined by observations.

Based on our analytical model, the leading/trailing asymmetry amplitude of the crater density that is generated by the MBAs is estimated to be $A_1^{\rm{c}}$ = 0.101--0.159,
while that generated by the NEOs is $A_1^{\rm{c}}$ = 0.117, which is equivalent to $N_{\rm{c}}({\rm{apex}}) / N_{\rm{c}}({\rm{antapex}})$ = 1.27. We easily but precisely reproduced the $N_{\rm{c}}$ distribution and the variation of $N_{\rm{c}}({\rm{apex}}) / N_{\rm{c}}({\rm{antapex}})$ as a function of $a_{\rm{M}}$ derived by \citet{LeFeuvre2011}, with $\alpha_{\rm{p}}$ = 2.6, $A_1^{\rm{c}}$ = 0.156, and $A_2^{\rm{c}}$ = 0.093. Our analytical model is applicable to other planet-satellite systems as long as $v_{\rm{enc}} > v_{\rm{orb}}$ and will be generalized to the case $v_{\rm{enc}} < v_{\rm{orb}}$ in future.

\begin{acknowledgements}
We thank the referee Bill Bottke for constructive and inspiring comments.
This research has been supported by the Key Development Program of Basic Research of China (Nos. 2013CB834900), the National Natural Science Foundations of China (Nos. 11003010 and 11333002), the Strategic Priority Research Program "The Emergence of Cosmological Structures" of the Chinese Academy of Sciences (Grant No. XDB09000000), the Natural Science Foundation for the Youth of Jiangsu Province (No. BK20130547) and the 985 project of Nanjing University and Superiority Discipline Construction Project of Jiangsu Province.
\end{acknowledgements}

\bibliographystyle{aa}
\bibliography{01}

\end{document}